\documentclass[a4paper,11pt]{article}
\pdfoutput=1
\usepackage{graphicx}
\usepackage{epsf,amsmath,bbold,amsfonts,stmaryrd}

\usepackage[utf8]{inputenc}
\usepackage{mathrsfs}
\usepackage{appendix}
\usepackage{amssymb}
\usepackage{float}
\usepackage{color}
\usepackage{cite}
\usepackage{hyperref}
\hypersetup{pageanchor=false}
\usepackage{indentfirst}
\usepackage{url}
\usepackage{float}
\usepackage{caption}
\usepackage[numbers,square,comma,sort&compress,merge]{natbib}

\def\nn{\nonumber}

\def\be{\begin{equation}}
\def\ee{\end{equation}}

\def\bea{\begin{eqnarray}}
\def\eea{\end{eqnarray}}

\def\ba{\begin{array}}
\def\ea{\end{array}}

\def\bc{\begin{center}}
\def\ec{\end{center}}

\def\bl{\begin{flushleft}}
\def\el{\end{flushleft}}

\def\br{\begin{flushright}}
\def\er{\end{flushright}}

\def\bi{\begin{itemize}}
\def\ei{\end{itemize}}

\def\bt{\begin{tabular}}
\def\et{\end{tabular}}

\numberwithin{equation}{section}

\begin{document}
\title{\textbf{Shadow of a Spinning Black Hole in an Expanding Universe}}
\author{Peng-Cheng Li$^{1,2}$,
Minyong Guo$^{1*}$,
and
Bin Chen$^{1,2,3}$}
\date{}

\maketitle

\vspace{-10mm}

\begin{center}
{\it
$^1$Center for High Energy Physics, Peking University,
No.5 Yiheyuan Rd, Beijing 100871, P. R. China\\\vspace{1mm}

$^2$Department of Physics and State Key Laboratory of Nuclear
Physics and Technology, Peking University, No.5 Yiheyuan Rd, Beijing
100871, P.R. China\\\vspace{1mm}

$^3$ Collaborative Innovation Center of Quantum Matter,
No.5 Yiheyuan Rd, Beijing 100871, P. R. China
}
\end{center}

\vspace{8mm}

\begin{abstract}
We study the influence of the cosmic expansion on the size of the shadow of a spinning black hole observed by a comoving observer. We first consider that the expansion is driven by a cosmological constant only and build the connection between the Kerr-de Sitter metric and the FLRW metric. We clarify that the notion of a comoving observer is well defined in the spacetime of a spinning black hole only in the sense of being asymptotic. The angular size of the shadow for a comoving observer is calculated. Significantly we find that the angular size approaches a non-zero finite value for a distant comoving observer, while it vanishes for a distant static observer. Furthermore, by adopting the approximate method proposed in \cite{Bisnovatyi-Kogan:2018vxl} we extend the study to the general  multi-component universe case. The results show that the difference between the horizontal and vertical angular size changes a lot, while their ratio, i.e. the oblateness, keeps unchanged when the supermassive spinning black hole is at a high redshift, due to the common amplification factor exerted  by the cosmic expansion. In addition, when $a=0$, our results agree with the previous studies in \cite{Perlick:2018iye,Bisnovatyi-Kogan:2018vxl}.
\end{abstract}

\vfill{\footnotesize Email: lipch2019@pku.edu.cn,\,\,minyongguo@pku.edu.cn,\,bchen01@pku.edu.cn,\\$~~~~~~*$ Corresponding author.}

\maketitle

\newpage
\section{Introduction}

In recent years, various kinds of astronomical observations strongly reveal that black holes do exist in our universe. The evidences so far are not limited to gravitational wave detections by LIGO and Virgo, but also include the images of black holes by Event Horizon Tecescope (EHT) \cite{Akiyama:2019cqa, Akiyama:2019brx,Akiyama:2019sww,Akiyama:2019bqs,Akiyama:2019fyp,Akiyama:2019eap}. However, compared to the precision of LIGO and Virgo, the pictures of black holes captured by EHT are not clear enough. One way to improve the accuracy and sharpness of the image is to vastly enhance the resolution of EHT by developing the optical interference technology, telescope technology and the like. On the other hand, it is also necessary to study the possible factors that may influence the shadow of a black hole in our universe.

The theoretical calculations of the black hole shadow began with Synge \cite{Synge:1966okc} who originally introduced the concept of \lq\lq{}escape cone\rq\rq{} which is known as the complement of the shadow of Schwarzschild black hole nowadays. He pointed out the edge of the shadow is rounded and also gave the formula of the angular radius for a static observer at infinity. In the following creative paper \cite{Bardeen:1973}, Bardeen extended the concept of black hole shadow to spinning black holes, and he found that the shape of the photon ring of the Kerr black hole could be deformed when the rotating parameter $a$ is not vanishing (the result can also be found in Chandrasekhar’s book \cite{Chandrasekhar:1992}). In particular, a portion of the photon ring\rq{}s edge becomes a vertical line segment, which is called NHEK-line for (near) extremal Kerr black hole. Straight after this work, Bardeen and Cunningham took an indepth study on the optical appearance of a star orbiting an extremal Kerr black hole by concentrating on the apparent position and the energy flux of the brightest images as seen by distant observers \cite{Cunningham:1972,Cunningham:1973} \footnote{For recent developments on this aspect, one can see \cite{Porfyriadis:2016gwb,Gralla:2017ufe, Guo:2018kis, Gates:2018hub, Long:2018tij,Yan:2019etp, Igata:2019pgb, Kapec:2019hro, Igata:2019hkz,Guo:2019lur,Guo:2019pte}.}. Along these lines, a number of works have been made to enrich our understanding of black hole shadows in general relativity and the modified gravity theory over the past four decades, see the examples in \cite{Grenzebach:2014fha,Amir:2016cen,Grenzebach:2015oea,Abdujabbarov:2016hnw,Dastan:2016vhb,Younsi:2016azx,Wang:2017hjl, Cunha:2018acu, Wang:2018eui, Hennigar:2018hza,Ovgun:2018tua,Haroon:2018ryd, Wang:2018prk,Wei:2019pjf, Kumar:2019ohr, Shaikh:2019fpu,Bambi:2019tjh,Konoplya:2019sns,Contreras:2019nih,Jusufi:2019nrn,Vagnozzi:2019apd,Zhu:2019ura, Ovgun:2019jdo, Contreras:2019cmf,Konoplya:2019goy,Konoplya:2019fpy, Das:2019sty, Lu:2019zxb,Chang:2019vni, Feng:2019zzn, Kumar:2019pjp, Ma:2019ybz, Kumar:2020hgm,Allahyari:2019jqz,Chang:2020miq}. In addition, we would like to stress that the key point in the calculations of the shadow among these works is to define a static observer in a static spacetime or a local rest static observer in a stationary spacetime. In other words, one cannot directly apply the standard methods proposed by Synge and Bardeen to a dynamical black hole or a black hole embedded in a time-dependent spacetime.

However, on the one hand,  black holes are generically not always eternal, since the matters keep falling into the black holes. It is   interesting to investigate the shadow in the process that matters collapse to form a black hole. On the other hand, it has been found that the universe we live is constantly expanding. These facts compel us to study the shadows beyond static or stationary black holes.  Akash Mishra \cite{Mishra:2019trb} has tried to understand the photon sphere and the black hole shadow in dynamically evolving spacetimes, including the
Vaidya, the RN-Vaidya and the de-Sitter Vaidya spacetimes, as well as the slowly rotating Kerr-Vaidya spacetime. Regarding a black hole embedded in an expanding universe, the authors in \cite{Perlick:2018iye} heuristically studied the Schwarzschild black hole shadow in an expanding universe only driven by a positive cosmological constant analytically. Their work has been extended to the multi-component universe  in \cite{Bisnovatyi-Kogan:2018vxl} by using an approximation method and in \cite{Tsupko:2019mfo} by using an analytic method. Based on these works, it was proposed that  the shadows of high-redshift black holes may be useful  in the cosmological studies \cite{Tsupko:2019pzg, Qi:2019zdk, Vagnozzi:2020quf}. All these studies have been focused on the spherical black holes without rotation. But almost all the black holes in our universe are spinning, and should be described by the Kerr spacetime and its cousins.  Thus, it is necessary and important to study the shadow of the Kerr black hole in an expanding universe in order to have  a better match with the images taken by the EHT in the future.

With such a purpose, in this paper we would like to study the effect of the expansion of our universe on the shadow of a spinning black hole. We begin our study on the expanding universe driven by a positive cosmological constant. In this case, we find  an exact form of the shadow  of the Kerr-dS black hole seen by a comoving observer following the trick in \cite{Perlick:2018iye}. Strictly speaking, the notion of a comoving observer in the FLRW universe is not well-defined in the presence of a spinning black hole, since the spinning locally breaks the isotropy of the universe.  In other words, the Kerr-dS spacetime cannot be foliated into spacelike slices which are conformally flat. Nevertheless, the comoving observer is approximately well-defined at the large distance away from the black hole. This fact allows us to use the following treatment: we first consider the shadow of Kerr-dS black hole observed by a locally static observer and then calculate the angular radius of the shadow of the Kerr black hole for the comoving observer with the cosmic expansion by using the standard aberration formula. We find that even though the black hole size seen by the static observer  tends to vanishing, the size seen by the comoving observer is not vanishing,  due to the magnification effect of the cosmic expansion.

Next we consider the multi-component FLRW universe and  investigate the shadow of a Kerr black hole in it. We apply the approximate method proposed in \cite{Bisnovatyi-Kogan:2018vxl}.  In this case, based on the size of the Kerr black hole shadow observed by a large distance local rest static observer, the effective proper size of the shadow can be found using the angular diameter redshift relation. In our work, we focus on the Kerr black hole, and discuss the influence of the rotating parameter $a$ and different inclination angles between the observer and the rotation axis of the black hole on the shadow. %And of course, our results are valid for Schwarzschild black holes when taking $a=0$.

The paper is organized as follows. In section \ref{section2}, we give a short review of calculating the shadow in the Kerr-dS spacetime as seen by a locally static observer, and introduce some characteristic angular diameters which are essential in the discussion. This is a necessary preparation for the following sections. In section \ref{section3}, we move to the core of this article, namely, the  shadow in the Kerr-dS spacetime as seen by a comoving observer. In section \ref{section4}, we explore the shadow of the Kerr black hole embedded in an expanding universe driven by different components. We give a summary in section \ref{summary}.

In this work, we have set the fundamental constants $c$ and $G$ to unity, and we will work in the convention $(-, +, +, +)$.

\section{Shadow in the Kerr-dS spacetime as seen by a local observer}\label{section2}

In this section we first give a review on the computations of the shadow of a Kerr-dS black hole observed by a locally non-rotating observer. The discussion here is mainly based on the instructive paper \cite{Grenzebach:2014fha}, but we also slightly add some supplements and present the necessary details for completeness and future calculations.

In terms of the Boyer-Lindquist(BL) coordinates $(t, r, \theta, \phi)$, the Kerr-dS metric is
written in the following compact form \cite{Akcay:2010vt}
\begin{eqnarray}\label{Kerr-dSmetric}
  d s^2  =  - \frac{\Delta_r}{\Sigma} \left[ \frac{d t}{\Xi \text{}} - a
  \sin^2 \theta \frac{d \phi}{\Xi \text{}} \right]^2 + \frac{\Sigma}{\Delta_r}
  d r^2 + \frac{\Sigma}{\Delta_{\theta}} d \theta^2 + \frac{\Delta_{\theta}
  \sin^2 \theta}{\Sigma} \left[ \frac{a d t}{\Xi \text{}} - (r^2 + a^2)
  \frac{d \phi}{\Xi \text{}} \right]^2,
\end{eqnarray}
where
\begin{eqnarray}
  \Sigma & = & r^2 + a^2 \cos^2 \theta, \\
  \Delta_r & = & (r^2 + a^2) \left( 1 - \frac{\Lambda}{3} r^2 \right) - 2 m r,  \label{Deltar}\\
  \Delta_{\theta} & = & 1 + \frac{a^2 \Lambda}{3} \cos^2 \theta, \\
  \Xi & = & 1 + \frac{a^2 \Lambda}{3} .
\end{eqnarray}
Note that $\Delta_r$ is a fourth-order polynomial of $r$, so formally we can
write it as
\begin{equation}
 \Delta_r= (r - r_+) (r - r_-) (r - r_C) \left( r - r_{- \, -} \right) .
\end{equation}
In general, the four roots of $\Delta_r=0$ are not all necessarily real. But in this paper
we only consider the case that the parameters $(\Lambda, m, a)$ take appropriate values
such that the  equation $\Delta_r=0$ yields four real roots. Here $r_{\pm}$ are the
outer and inner black hole horizons, $r_C$ is the cosmological horizon and $r_{-\,-}< 0_{}$
is the other cosmological horizon ``inside'' the singularity. To satisfy the
condition we just imposed,  the bound on the value of $\Lambda$ is easily obtained as $0<m^2 \Lambda <1/9\simeq 0.11$,
 when $a=0$. For simplicity and without loss of generality, we choose the angular momentum of the black hole to be non-negative, that is to say, we always have $a\geq0$. If $a>0$ , then the upper bound of $m^2\Lambda$ is
slightly larger than $0.11$. In turn, the presence of $\Lambda$ enlarges the
range of $a$ such that the upper bound of $a$ can be slightly larger than $m$.
For example, when $\Lambda$ is very small,  which is close to the reality world, we can easily obtain the four real roots analytically up to the next-to-leading order in $\Lambda$,
that is
\be\label{rrc}
r_C=\sqrt{\frac{3}{\Lambda}}-1,\quad r_{- \, -}=-\sqrt{\frac{3}{\Lambda}}-1,
\ee
\be
r_\pm=m\pm\sqrt{m^2-a^2}+\frac{\Lambda}{3} m \left(4 m^2 \left(1\pm\frac{m}{\sqrt{m^2-a^2}}\right)
-a^2 \left(1\pm\frac{3 m}{\sqrt{m^2-a^2}}\right)\right).
\ee
Note that the expressions of $r_\pm$ does not allow us to find the upper bound of the rotation parameter, namely the analogue of the Kerr bound.
This is because when deriving $r_\pm$, $a$ was already assumed to be independent of $\Lambda$, so the upper bound is just $a\leq m$. In fact,  the upper bound  of $a$ up  to the linear order of $\Lambda$ can be read directly from
(\ref{Deltar}):  $a_{max}=m+m^3\Lambda/3$.

In the region between the outer black hole horizon $r_+$ and the cosmological horizon $r_C$, one finds the vector $\partial_r$ is spacelike and $\partial_t$ is the timelike Killing vector, thus this
region is the so-called {\em domain of outer communication}. This region guarantees the causality of the spacetime, thus a static observer beyond this region is not well-defined. However, for
comoving observers with respect to the cosmic expansion, this is not true. Even after crossing the cosmological horizon \footnote{We are thankful to anonymous referee for pointing out that this is event horizon which differs from the particle horizon. }, a
comoving observer can still receive light signals from the {\em domain of outer communication}.

The null geodesic equations are completely integrable, since it admits four constants along the motion of each photon:

(1) the mass $ g_{\mu \nu}  p^{\mu}
p^{\nu}=0$,

(2) the total energy $E = - p \cdot \partial_t$,

(3) the angular momentum $L = p \cdot \partial_{\phi}$, and

(4) the Carter constant $Q$, respectively.

With the help of these constants along the motion, the geodesic equations can be written
in the first-order form \cite{Grenzebach:2014fha}
\begin{eqnarray}
  \Sigma \dot{t} & = & \frac{a \sin^2 \theta (L - a E \sin^2
  \theta)}{\Delta_{\theta} \sin^2 \theta} + \frac{(\Sigma + a^2 \sin^2 \theta)
  ((\Sigma + a^2 \sin^2 \theta) E - a L)}{\Delta_r}, \\
  \Sigma \dot{\phi} & = & \frac{L - a \sin^2 \theta}{\Delta_{\theta} \sin^2
  \theta} + \frac{a (\Sigma + a^2 \sin^2 \theta) E - a L}{\Delta_r}, \\
  \Sigma^2  \dot{\theta}^2 & = & \Delta_{\theta}  (Q + (L - a E)^2) - \frac{(a
  \sin^2 \theta E - L)^2}{\sin^2 \theta},\label{thetadot} \\
  \Sigma^2  \dot{r}^2 & = & ((\Sigma + a^2 \sin^2 \theta) E - a L)^2 -
  \Delta_r  (Q + (L - a E)^2) \nn\\
  &\equiv& \mathcal{R} (r),
\end{eqnarray}
where we have defined a radial function $\mathcal{R} (r)$.
In the Kerr-dS spacetime there are unstable spherical photon orbits with $r
=$constant. All such orbits occupy a region which is known as the photon region. Since the
photon orbits are independent of the energy, we introduce the following
dimensionless quantities to characterize the photon orbits
\begin{equation}
  \xi = \frac{L}{E}, \quad \eta = \frac{Q}{E^2} .
\end{equation}
The spherical orbits occur when $\mathcal{R} = \dot{\mathcal{R}} = 0$, then we can
express the above two rescaled quantities in terms of  the radius of the orbit,
\begin{eqnarray}
  \xi & = & \frac{r^2 + a^2}{a} - \frac{4 r \Delta_r}{a \Delta_r'}, \label{xi}\\
  \eta & = & \frac{16 r^2 \Delta_r}{(\Delta_r')^2} - \left( \frac{r^2}{a} -
  \frac{4 r \Delta_r}{a \Delta_r'} \right)^2 .\label{eta}
\end{eqnarray}
Plugging these expressions into (\ref{thetadot}), we find an inequality that determines the photon region
\be(4 r \Delta_r - \Sigma \Delta_r')^2 \leq 16 a^2 r^2 \Delta_r
   \Delta_{\theta} \sin^2 \theta .\label{inequality}
\ee
For $\Lambda = 0$, the radial range of the photon region
is determined by the roots of $\eta = 0$,
\be
r \in [r_{p -}, r_{p +}], \ee
with
\be
  r_{p \pm}=2 m \left( 1 + \cos \varphi_\pm \right),\quad \varphi_\pm=\frac{2}{3}\cos^{-1}\left(\pm\frac{a}{m}\right).\label{rpp}
\ee
The presence of a non-vanishing $\Lambda$ will slightly change the range of the photon region. In this case, the
inequality \eqref{inequality} is still a third order polynomial of $r$, so it can be solved
analytically. For example, let us consider the small $\Lambda$ case, then
up to the linear order of $\Lambda$, the corrections to $r_{p\pm}$  are given by
\be
\delta r_{p \pm}  = -2am \Lambda\left[a+3a \cos\varphi_\pm+\frac{(27m^2-25a^2)\sin\varphi_\pm}{9\sqrt{m^2-a^2}}\right].\label{drpp}
\ee
In the non-rotating case $a = 0$, the above inequality degenerates
into an equality,
\be
4 r \Delta_r = \Sigma\, \Delta_r',
\ee
and the photon region degenerates into a photon sphere
with $r_p = 3 m$.   This result is independent of $\Lambda$.

Next, we move to the calculation of the critical curve (which is also known as the contour of the black hole shadow, or vaguely called the photon ring) of the Kerr-dS black hole as seen by a locally static observer. First of all, we assume that the observer is located in the  frame in the following form
\begin{eqnarray}\label{et}
  \hat{e}_{(t)} & = & \sqrt{\frac{g_{\phi \phi}}{g_{t \phi}^2 - g_{t t} g_{\phi
  \phi}}} \left( \partial_t - \frac{g_{t \phi}}{g_{\phi \phi}} \partial_{\phi}
  \right),\label{et}\\
  \hat{e}_{(r)} & = & \frac{1}{\sqrt{g_{r r}}} \partial_r,\\
  \hat{e}_{(\theta)} & = & \frac{1}{\sqrt{g_{\theta \theta}}} \partial_{\theta},\\
  \hat{e}_{(\phi)} & = & \frac{1}{\sqrt{g_{\phi \phi}}} \partial_{\phi},
\end{eqnarray}
where $\hat{e}_{(t)}$ is timelike in the {\em domain of outer communication} and the
rest bases are spacelike vectors. It is easy to check that these bases are normalized
and orthogonal to each other. We can assign the four-velocity of the
observer as $\hat{e}_{(t)}$, that is, $\hat{u}\equiv \hat{e}_{(t)}$, in this case the observer is locally static in the
given frame. Moreover, since $\hat{e}_{(t)} \cdot \partial_{\phi} = 0$ the observer in this local rest frame has zero angular momentum with respect to infinity.
Hence this frame is sometimes called the ZAMO reference frame, standing for
the zero angular momentum observers.

\begin{figure}[h]
\begin{center}
\includegraphics[width=105mm,angle=0]{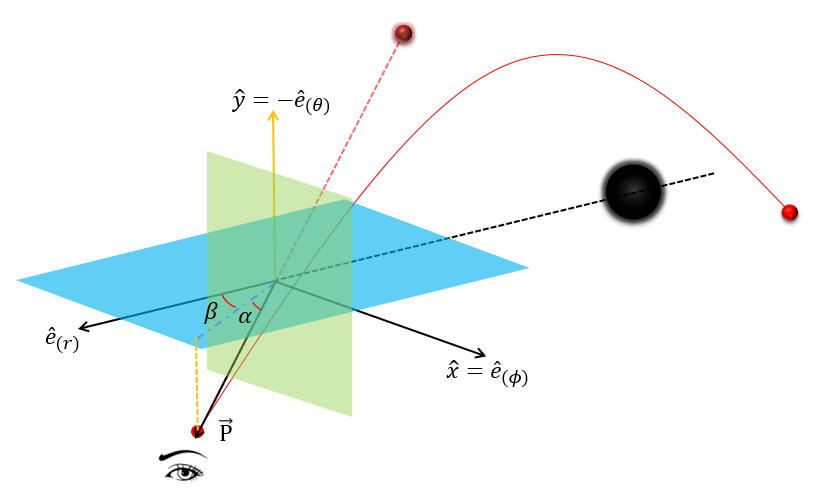}
\end{center}
\vspace{-5mm}
 \caption { The projection of the photon's momentum $\vec{P}$ in the observer's frame and the solid angles $(\alpha, \beta)$.  }\label{referenceframe}
\end{figure}

For the null geodesics, the four-momentum $p^{\mu}$ can be projected onto the
four bases of the observer's frame. This is  called a local measurement, giving the quantities measured by the observer
\begin{eqnarray}
  p^{(t)} & = & - p_{\mu} \hat{e}_{(t)}^{\mu}, \\
  p^{(i)} & = & p_{\mu} \hat{e}_{(i)}^{\mu}, \quad i = r, \theta, \phi.
\end{eqnarray}
On the other hand, since the photon is massless, in the observer's frame, one
has $| \vec{P} | = p^{(t)}$, and we can introduce the observation angles $(\alpha,
\beta)$, by \cite{Cunha:2016bpi}
\begin{eqnarray}
  p^{(r)} & = & | \vec{P} | \cos \alpha \cos \beta, \\
  p^{(\theta)} & = & | \vec{P} | \sin \alpha, \\
  p^{(\phi)} & = & | \vec{P} | \cos \alpha \sin \beta ,
\end{eqnarray}
as is shown in Fig. \ref{referenceframe}. From the geodesic equations we obtain
\begin{eqnarray}
  \sin \alpha& =&  \frac{p^{(\theta)}}{p^{(t)}}=\pm\frac{1}{\zeta-\gamma \xi}
  \sqrt{\frac{\Delta_\theta\sin^2\theta(\eta+(\xi-a)^2)-(a\sin^2\theta-\xi)^2}{\Sigma \Delta_\theta\sin^2\theta}}\Bigg|_{(r_O,\theta_O)} ,\label{angles1} \\
    \tan \beta &=& \frac{p^{(\phi)}}{p^{(r)}}=\frac{\xi \sqrt{\Sigma \Delta_r}}
    {\sqrt{g_{\phi\phi}}\sqrt{((\Sigma+a^2\sin^2\theta)-a\xi)^2-\Delta_r(\eta+(\xi-a)^2)}}\Bigg|_{(r_O,\theta_O)},\label{angles2}
\end{eqnarray}
where  we have used the abbreviations $\zeta= \hat{e}_{(t)}^t$ and $\gamma=\hat{e}_{(t)}^\phi$ to simplify the expression.
Note that in the expressions, only $(\xi, \eta)$ depend on the radius of the photon region, on the other hand, all $(r, \theta)$ take values at the position of the
observer, i.e. $(r_O, \theta_O)$, where $\theta_O$  is the inclination angle between the observer and the direction of the rotation axis of the black hole.

Furthermore, we can introduce the Cartesian coordinate $(x, y)$ for the apparent position on the plane of the sky of the observer by \footnote{In the literatures \cite{Bardeen:1973, Chandrasekhar:1992}, the Cartesian coordinates are defined by
$x \equiv - r_O \cos \alpha \sin\beta$, $y \equiv r_O \sin \alpha $. Our definition is similar to the one in \cite{Cunha:2016bpi}.} %We find different definitions have no significant effect on the later discussions. }
\begin{equation}
  x \equiv - r_O \beta, \quad y \equiv r_O \alpha .\label{xy}
\end{equation}

\begin{figure}[h]
\begin{centering}
\includegraphics[width=160mm,angle=0]{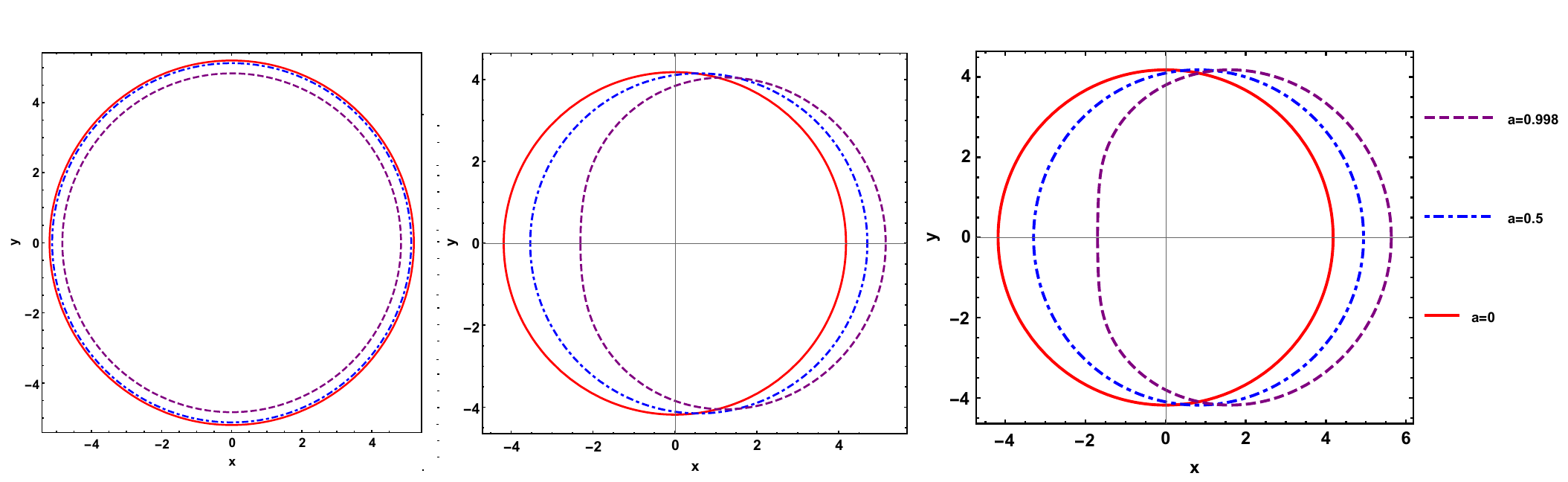}
\par\end{centering}
\caption{Examples of shadows of the Kerr-dS black hole for different rotation parameters and inclination angles. From left panel to the right panel, the inclination angle $\theta_O$ takes $0$, $\pi/4$ and $\pi/2$, respectively.}\label{variousangles}
\end{figure}

\begin{figure}[h]
\begin{center}
\includegraphics[width=75mm,angle=0]{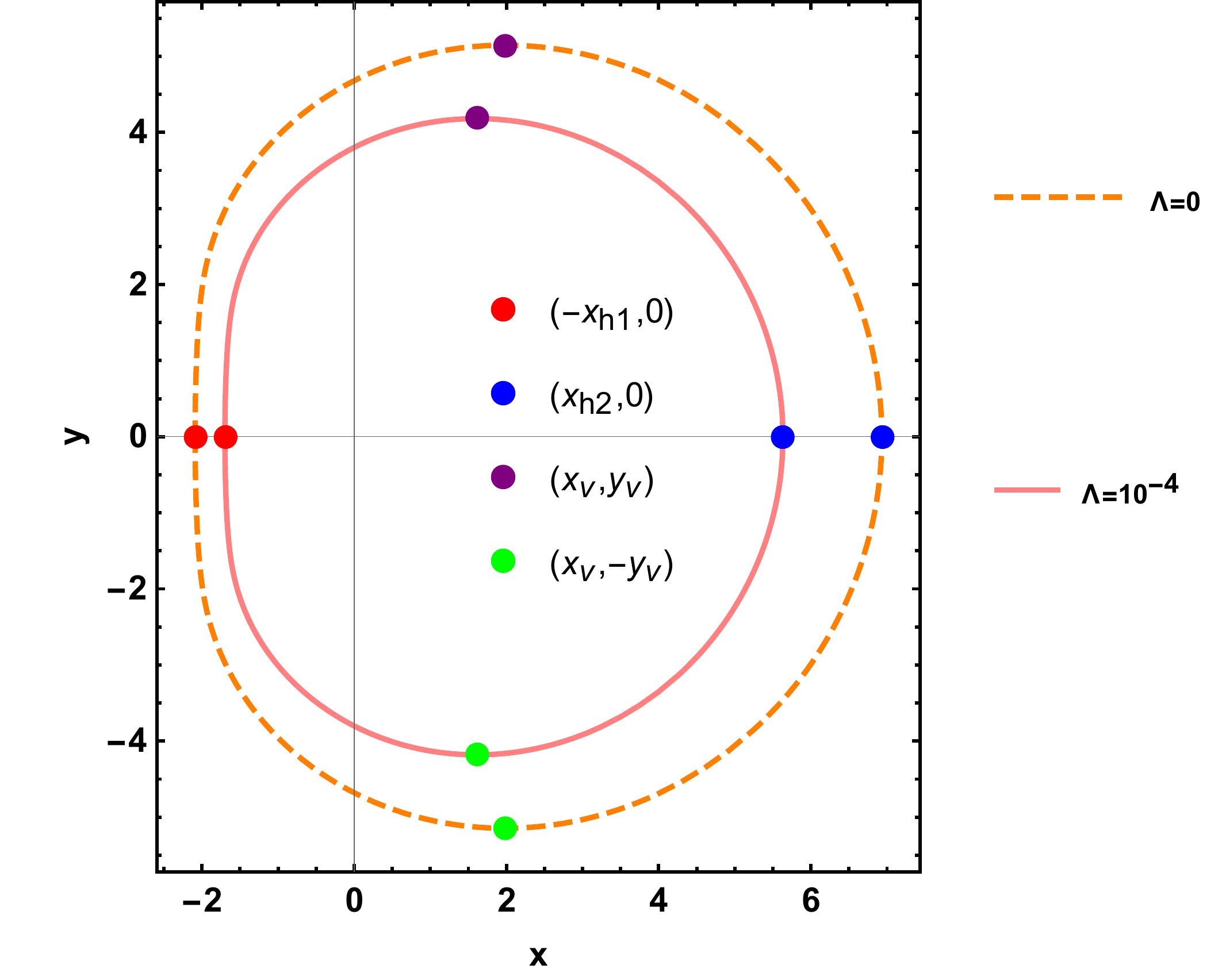}
\end{center}
\vspace{-5mm}
 \caption {Shadows of the Kerr-dS black hole as seen by a locally static observer with different values of the cosmological constant. The dots on the $x$ axis denote the points at which $\alpha=0$ and the dots along the vertical direction correspond to the position at which $\alpha$ takes maximum value.  }\label{ShdaowKerrdS}
\end{figure}
The boundary curve of the shadow corresponds to the null geodesics \ that
asymptotically approach the photon region. So these null geodesics are
characterized by the two constants of motion $(\xi, \eta)$, whose  relations
with the photon region are given by Eqs. (\ref{xi}) and (\ref{eta}). Inserting Eqs. (\ref{xi}) and (\ref{eta}) into
Eq. (\ref{xy}), we then obtain the boundary curve of the shadow on the
observer's sky parameterized with the radius of the photon region.

In Fig. \ref{variousangles} we show the shadows of the Kerr-dS black hole for several rotation parameters and inclination angles, where we fix $m=1, r_O=100$. Qualitatively, they share similar features with the shadow of the Kerr black hole. For example, with the increase of the angular momentum, the shape of the shadow gradually becomes non-circular. But for the observer staying at the north of the black hole, i.e. $\theta_O=0$, the shadow will always be circular. The difference  between the shadows of the spinning and the static black holes becomes the most significant when the spinning black hole is extremal and when the observer lies in the equator plane of the black hole.

In addition, in Fig. \ref{ShdaowKerrdS}, we show the effect of the cosmological constant on the black hole shadow, where we take $\theta_O=\pi/2$ and $a=0.998$. One can observe the critical curve becomes smaller when the cosmological constant is not vanishing. To explain this, we consider that the static observer is located far away from the black hole, i.e. $r_+\ll r_O \leq r_c$. To take such limit, the cosmological constant $\Lambda$ must be very small such that the cosmological horizon is  large enough, otherwise the locally static observer would exceed the cosmological horizon. In the next section we will restrict ourself to this particular case. From Eqs. (\ref{angles1}) and (\ref{angles2}), we obtain the asymptotic form of the observation angles
\begin{eqnarray}\label{sima}
  \sin \alpha& \to& \pm
  \frac{\sqrt{1-\frac{\Lambda}{3} r_O^2}\sqrt{\eta+(a-\xi )^2 -(\xi \csc \theta_O-a \sin \theta_O)^2}}{r_O},\\\label{simb}
    \tan \beta &\to& \frac{\xi \sqrt{1-\frac{\Lambda}{3} r_O^2}}
    { r_O \sin\theta_O}.
\end{eqnarray}
\begin{figure}[h]
\begin{center}
\includegraphics[width=90mm,angle=0]{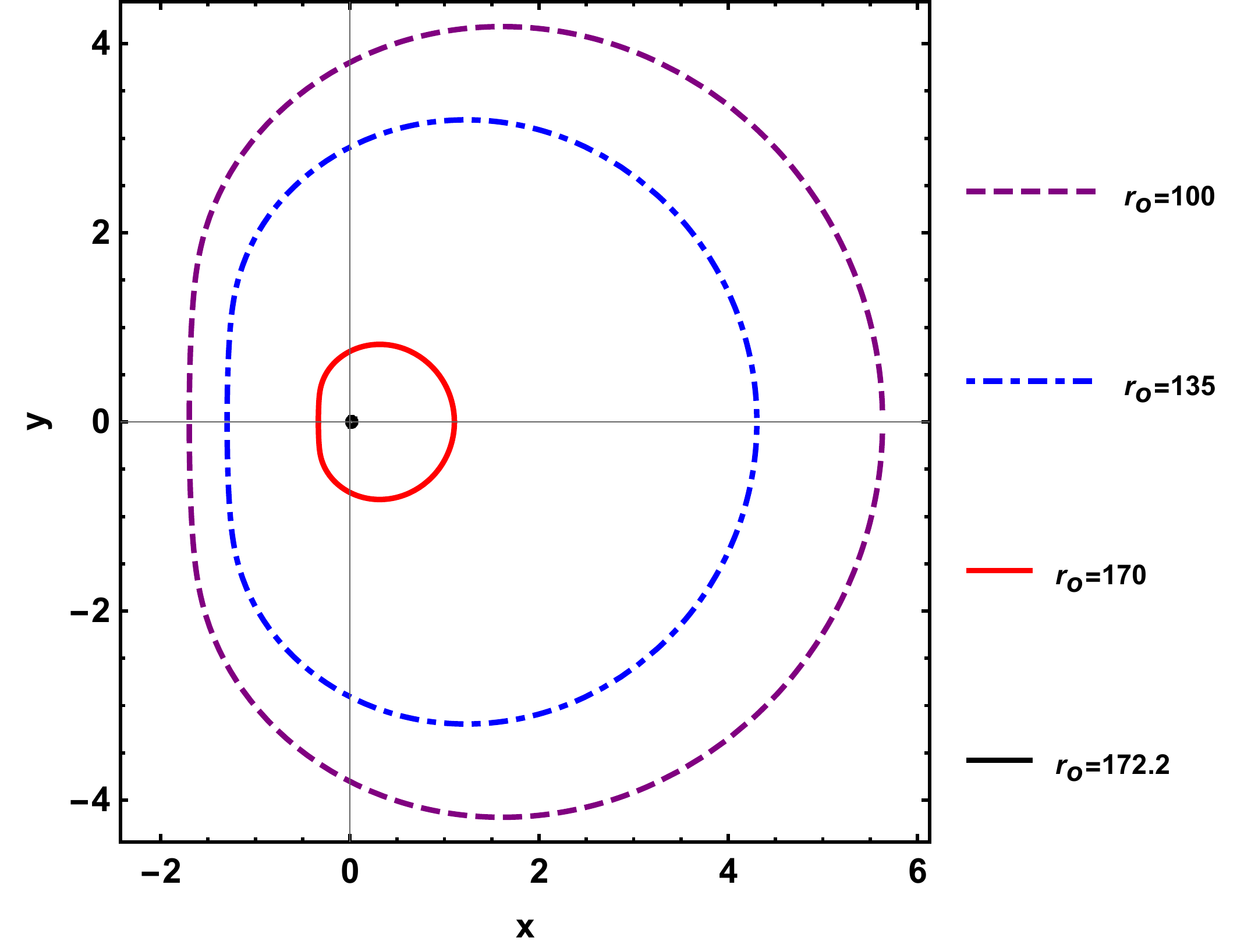}
\end{center}
\vspace{-5mm}
 \caption { The shadows vary with respect to the distance of the observer $r_O$. We have choosen $m=1, a=0.998, \Lambda=10^{-4}, \theta_O=\pi/2$, and the corresponding cosmological horizon is $r_c\simeq 172.2$. }\label{varyro}
\end{figure}
Combining with Eq. (\ref{xy}), we have
\be
x^2+y^2=(1-\frac{\Lambda}{3}r_O^2)(\xi^2+\eta+a^2\cos^2\theta_O)
\ee
When the observer is located at $(r_O, \theta_O=\frac{\pi}{2})$ and if we pay our attention to the edge of the critical curve on the $x$ axis, the above equation gives
\be
(x^2+y^2)|_{r=r_{p\pm}+\delta r_{p \pm}}=(1-\frac{\Lambda}{3}r_O^2)\xi^2|_{r=r_{p\pm}+\delta r_{p \pm}}=(1-\frac{\Lambda}{3}r_O^2)\xi^2|_{r=r_{p\pm}}+\mathcal{O}(\Lambda^2).
\ee
This explains why the size of shadow becomes smaller when the cosmological constant is nonvanishing.

Moreover, from Eqs. (\ref{sima}) and (\ref{simb}), we may conclude that when the observer is approaching the cosmological horizon $r_O\to r_c\simeq\sqrt{3/\Lambda}$,
\begin{eqnarray}
 r_O \sin \alpha \to0,\quad  r_O  \tan \beta \to 0.
\end{eqnarray}
 In Fig. \ref{varyro}, we show the shadow sizes versus the distance of the observer. In contrast, as we will see in the next section, the size of the black hole shadow as seen by an observer comoving with the cosmic expansion is finite even the observer is far away from the black hole.
This interesting phenomenon was first found in \cite{Perlick:2018iye} for the Schwarzschild
black hole.
\begin{figure}[h]
\begin{center}
\includegraphics[width=65mm,angle=0]{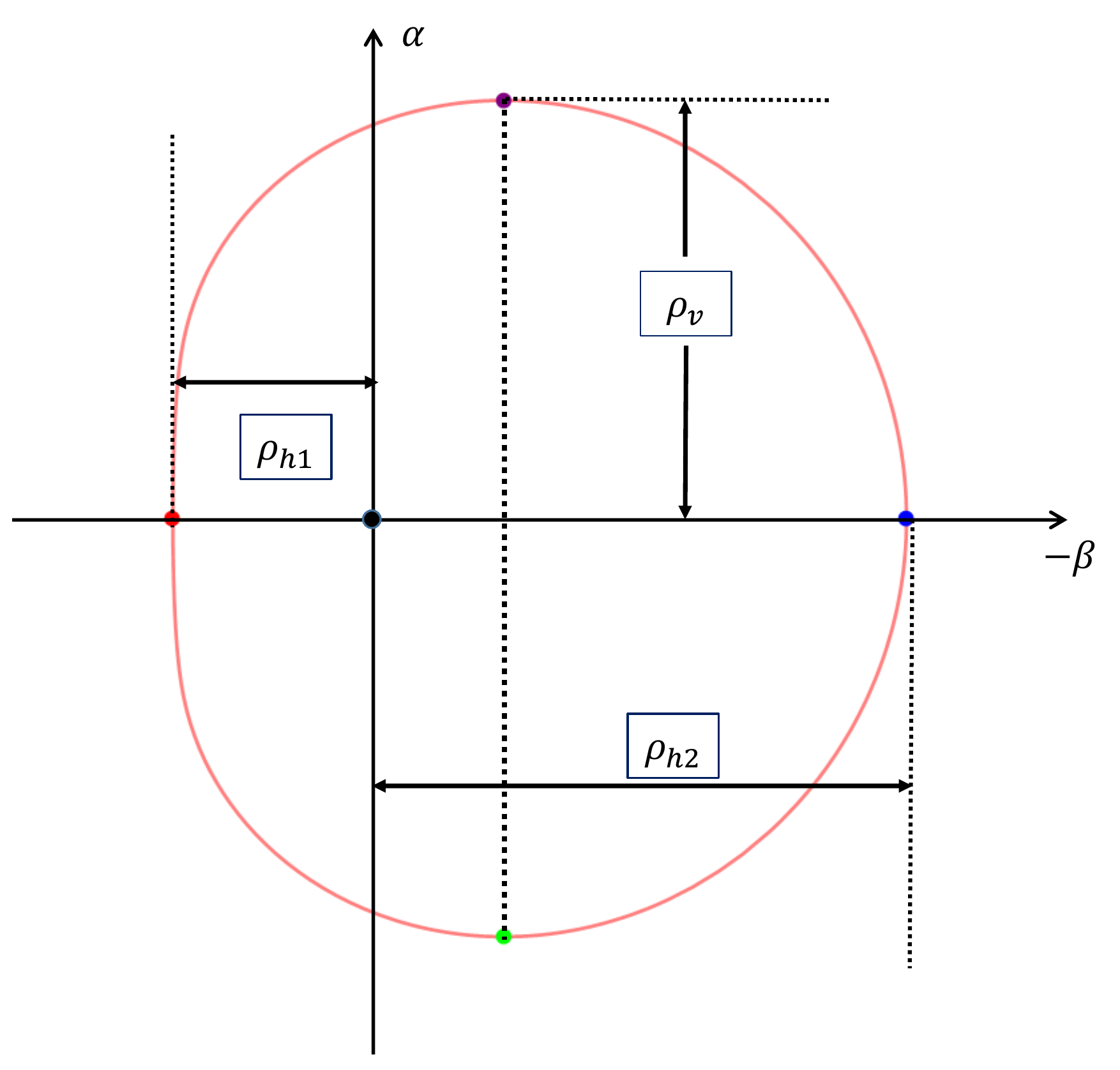}
\end{center}
\vspace{-5mm}
 \caption { Angular radii of the shadow of a spinning black hole. The horizontal axis is $-\beta$ and the vertical axis
 is $\alpha$.   }\label{angleradii}
\end{figure}

In the non-rotational case, the silhouette of the shadow is a circle.  However in the presence of the rotation, the shape is no longer
a circle, so it cannot be trivially characterized by a single angular radius. In
this case, we follow the strategy used in \cite{Grenzebach:2015oea} to characterize the size
of the shadow in terms of the horizontal and vertical angular diameters of the
shadow. From the analytical formulas for the boundary curve of the shadow we
can find the expressions for the horizontal and vertical angular diameters of
the shadow, as shown in Fig. \ref{angleradii} and marked in Fig. \ref{ShdaowKerrdS}. Due to the symmetry, the angular diameters $\Delta h$ and $\Delta
v$ are determined by three radii $\rho_{h 1}$, $\rho_{h 2}$ and $\rho_v$,
\begin{eqnarray}
  \Delta h & = & \rho_{h 1} + \rho_{h 2}, \label{horizontalang}\\
  \Delta v & = & 2 \rho_v,\label{verticalang}
\end{eqnarray}
where $\rho_{h i}$, $i = 1, 2$ are the absolute values $-\beta$ takes when $\alpha =
0$, and $\rho_v$ is the maximum value $\alpha$ takes. Formally, we can denote the corresponding photon region radius as $r_{hi}$ and $r_{v}$.  In general, we find the angular radii
 of the shadow cannot be obtained in a closed form, because the corresponding values of the photon region radius are the zeros of a polynomial of higher than fourth order. So the angular radii
 of the shadow have to be found numerically.

However, in some particular cases, it is feasible to get the analytical formulas for the angular radii. For example, for the observer in the
equatorial plane, i.e. $\theta_O=\pi/2$, $\alpha=0$ is equivalent to
  \be\label{Dethorizonangles}
  (4 r \Delta_r - r^2 \Delta_r')^2= 16 a^2 r^2 \Delta_r.
\ee
The solutions to this equation are just the bounds of the photon regions in the radial direction, which for the small $\Lambda$ case are simply given by (\ref{rpp}) and (\ref{drpp}), i.e. $r_{h1}=r_{p-}+\delta r_{p-}$ and $r_{h2}=r_{p+}+\delta r_{p+}$. Plugging these into Eqs. (\ref{xi}) and (\ref{eta}), and further into  Eqs. (\ref{angles1}) and (\ref{angles2}) we then obtain the angular radii of the shadow as functions of the observer's position $r_O$.

Another particular case is the non-rotating limit $a=0$, in which case the solution to the equation (\ref{Dethorizonangles}) is just $r_{hi} = 3 m$ and the horizontal angular radii of the
shadow are given by
\be \label{shadowSchdS}
\sin^2 \rho_{h 1} = \sin^2 \rho_{h 2} =\frac{9 m^2 \left(6 m+\Lambda  r_O^3-3r_O\right)}{r_O^3 \left(9 \Lambda  m^2-1\right)}.
\ee
This reproduces the result found by Stuchlik and Hledik \cite{Stuchlik:1999qk} and when $\Lambda=0$ it reduces to the well-known result calculated by Synge for the shadow of a Schwarzschild black
hole \cite{Synge:1966okc}. Moreover, in this case the angular size of the shadow for static observers at large distance, i.e. $r_O\gg m$, is simplified to
\be
\sin^2 \rho_{h 1} = \sin^2 \rho_{h 2} \simeq\frac{27 m^2}{r_O^2 }
 \left(1-\frac{\Lambda  r_O^2}{3}\right),
\ee
which is the same as the result in \cite{Perlick:2018iye}. As we mentioned before, the large distance limit is taken under the condition $\Lambda m^2 \ll 1$.

On the other hand, $\alpha$ takes a maximum at
\be
 \left( \frac{p^{(\theta)}}{p^{(t)}} \right)' = 0,
 \ee
 where the prime denotes the derivative with respect to the photon region radius. For $\theta_O=\pi/2$,
the solution to this equation is given by
\be\label{rofalphamax}
r_v = \frac{9 m r_O \left(a^2+r_O^2\right)}{a^4 \Lambda  r_O+a^2 \left(12 m+\Lambda  r_O^3+3 r_O\right)+3r_O^3}.
\ee
The explicit expression for $\sin \rho_v$ is still complicated so we shall not show it here. But in the non-rotating limit,  the result simplifies to be the same as (\ref{shadowSchdS}), as expected.

\section{Shadow in the Kerr-dS spacetime as seen by a comoving observer}\label{section3}

In this section, we would like to investigate the influence of a cosmic expansion on the shadow of the spinning
black hole. We will study the shadow as seen by a comoving observer. Our treatment is inspired by the one in  \cite{Perlick:2018iye} in which the shadow of  non-spinning black holes was studied. In  \cite{Perlick:2018iye},  based on the connection between the Schwarschild-dS black hole and the McVittie metric \cite{McVittie:1933}, the black hole shadow as seen by a comoving observer with the cosmic expansion was analytically calculated. The McVittie metric describes a non-rotational black hole embedding in an expanding universe. If the cosmic expansion is driven by a cosmological constant only, then through a coordinate transformation, the Schwarschild-dS metric can be related to the McVittie metric and the black hole shadows for different observers are related as well. Here we generalize the study in \cite{Perlick:2018iye} to the rotating black hole case.

To our knowledge, there are several solutions describing a spinning black hole
embedded in an expanding universe, such as the one constructed by Vaidy \cite{Vaidya:1977zza},
the metric proposed by Thakurta \cite{Thakurta:1981}  and so on. All these solutions share the
same property: when the mass of the black hole is vanishing, the solution
reduces to a FLRW metric, and when the scale factor is equal to unity, the
solution become the Kerr metric.
If the expanding universe is totally driven by a positive cosmological constant, then the FLRW metric with vanishing curvature
is given by \footnote{In this work, we focus on the flat FLRW universe. }
\begin{equation}\label{FLRW}
  d s^2 = - d \bar{t}^2 + a^2 d\bar{r}^2 + a^2 \bar{r}^2 (d \bar{\theta}^2 + \sin^2 \bar{\theta} d \bar{\phi}^2),
\end{equation}
with
\begin{equation}
a(\bar{t}) = e^{H_0 \bar{t}},\quad H_0 = \sqrt{\frac{\Lambda}{3}}.
\end{equation}
Now we would like to study the relation between the Kerr-dS metric and the metric of the cosmological spinning black hole. The starting point, as we mentioned
before, is that the metric of the cosmological spinning black hole should reduce to (\ref{FLRW}) when the mass is
zero and reduce to the Kerr metric when $H_0=0$.
\subsection{Coordinate transformation}
It is well known that the metric of the de Sitter spacetime is given by
\begin{equation}
    d s^2 =  - \left( 1 - H_0^2 R^2 \right) d T^2 + \frac{d
    R^2}{\left( 1 - H_0^2 R^2 \right)} + R^2 d \Theta^2 + R^2
    \sin^2 \Theta d \Phi^2.
\end{equation}
Now we will show that the de Sitter metric can be related to the FLRW metric (\ref{FLRW}) via a simple
coordinate transformation. First, let us introduce
\begin{equation}
  T = \bar{t} + f (R),\quad
  f' (R) =\frac{H_0^{} R}{1 - H_0^2 R^2}.
\end{equation}
Plugging this into the  de Sitter metric leads to
\begin{equation}
  d s^2 = - (1 - H_0^2 R^2) d \bar{t}^2 + 2 H_0^{} R d \bar{t} d R + d R^2 + R^2 (d
  \Theta^2 + \sin^2 \Theta d \Phi^2) .
\end{equation}
Furthermore  introducing
\begin{equation}
  R = a (\bar{t})\, \bar{r},\quad \bar{\theta}=\Theta,\quad \bar{\phi}=\Phi,
\end{equation}
then one finds that the de Sitter metric reduces to  the standard FLRW metric (\ref{FLRW}).
Therefore, we find the coordinate transformation that relates the de Sitter metric with the FLRW metric, viz.,
\begin{eqnarray}
  d \bar{t} & = & d T - \frac{H_0^{} R}{1 - H_0^2 R^2} d R,\label{coorTrans1} \\
  e^{H_0 \bar{t}} d \bar{r} & = & -  H_0 R d T + \frac{1}{1 - H_0^2 R^2} d R.\label{coorTrans2}
\end{eqnarray}
Equivalently, the coordinate transformation can be rewritten in terms of the coordinate
basis vectors as
\begin{eqnarray}
  \partial_{\bar{t}} & = & \frac{1}{1 - H_0^2 R^2} \partial_T + H_0 R \partial_R,\label{ptbar} \\
  e^{- H_0 \bar{t}} \partial_{\bar{r}} & = & \frac{H_0 R}{1 - H_0^2 R^2} \partial_T +
  \partial_R .
\end{eqnarray}
Next, we notice that the Kerr-dS metric (\ref{Kerr-dSmetric}) in the vanishing mass limit, i.e. $m=0$, becomes
\begin{eqnarray}\label{KerrdS_massless}
  d s^2 & = &(r^2 + a^2 \cos^2 \theta) \left[ \frac{d r^2}{(r^2 + a^2) \left( 1 -
  H_0^2r^2 \right)}  + \frac{d \theta^2}{1 +
  \frac{a^2 \Lambda}{3} \cos^2 \theta}  \right] \nonumber\\
  &  & + \frac{\left( 1 + a^2  H_0^2\cos^2 \theta \right) \sin^2
  \theta}{r^2 + a^2 \cos^2 \theta} \left[ \frac{a d t-(r^2 + a^2)d \phi}{1+a^2
  H_0^2}  \right]^2 \nonumber\\
  &  & - \frac{(r^2 + a^2) \left(1-H_0^2 r^2 \right)}{r^2 +
  a^2 \cos^2 \theta} \left[ \frac{d t- a
  \sin^2 \theta d \phi}{1+a^2 H_0^2}\right]^2 .
\end{eqnarray}
Actually, this is essentially the de Sitter metric, which is realized by performing the following coordinate transformations \cite{Akcay:2010vt}:
 \begin{eqnarray}
  d T & = & \frac{d t}{\Xi },\label{Tf_KerrdS_m_0a} \\
  \Phi & = & \phi - \frac{a t}{ \Xi\text{}}H_0^2,\label{Tf_KerrdS_m_0b} \\
  R \cos \Theta & = & r \cos \theta,\label{Tf_KerrdS_m_0c} \\
  R^2 & = & \frac{1}{\Xi \text{}^{} \text{}} [r^2 \Delta_{\theta} + a^2 \sin^2
  \theta].\label{Tf_KerrdS_m_0d}
\end{eqnarray}
Obviously, this transformation becomes trivial as $a=0$, since the BL coordinates become the ordinary Schwarzschild coordinates in this case. Moreover, if $\Lambda=0$, then this transformation relates the ellipsoidal coordinates with the
Schwarzschild coordinates. Thus, in the vanishing mass limit, the Kerr-dS metric is just the de Sitter metric and can be related to the FLRW metric through the coordinate
transformations (\ref{coorTrans1}) and (\ref{coorTrans2}).

The above discussion may help us to find the metric  describing a cosmological spinning black hole. Firstly, if the coordinate transformations (\ref{Tf_KerrdS_m_0a}), (\ref{Tf_KerrdS_m_0b}),
(\ref{Tf_KerrdS_m_0c}) and (\ref{Tf_KerrdS_m_0d}) are applied to the general Kerr-dS metric (\ref{Kerr-dSmetric}) instead of its massless limit (\ref{KerrdS_massless}), and the coordinate transformations (\ref{coorTrans1}) and (\ref{coorTrans2}) are carried out subsequently, then we would obtain a metric which in the massless limit reduces to the FLRW metric (\ref{FLRW}), and in the limit $\Lambda=0$ reduces to the Kerr metric in the Schwarzschild-like coordinate system.
This metric describes essentially a spinning black hole embedded in an expanding universe driven by a positive cosmological constant.% as the metric owns the basic features that a cosmological spinning black hole should have.

If $a=0$, the constructed metric describes a non-spinning cosmological black hole.  However, the reduced metric  is different from the solution found by McVittie in 1933 \cite{McVittie:1933}. The McVittie metric was obtained in the isotropic coordinates, which are naturally  the privileged comoving coordinates used to describe an isotropic universe. In these coordinates, the time coordinate basis vector can be assigned as the four-velocity of a comoving observer with respect to the cosmic expansion. It seems impossible to express the above constructed metric for the cosmological spinning black hole in similar isotropic coordinates, because the Kerr spacetime is a priori axisymmetric. Even if corotating with the Kerr black hole, the observer would not perceive an isotropic space. In other words, the universe cannot be foliated into spacelike slices such that each three-dimensional slice is conformally flat \cite{Garat:2000pn}. Therefore there is no comoving observer near the cosmological spinning black hole.

Nevertheless, as we will show in the next subsection the notion of the comoving observer can be well defined at least at large distance. In that case, the spacetime near the observer is not spinning any more, so the observer can be set to comoving with cosmic expansion. From a practical point of view, the observer on the earth is sufficiently far away from the massive black holes in the universe, so it seems reasonable to take the approximation that the comoving observer lies at large distance with respect to the observable black hole.

\subsection{Shadow for comoving observer at large distance}

Since the cosmological black hole is derived from the original Kerr-dS black hole via some coordinate transformations, different coordinate systems represent different reference frames. An event that occurs in the spacetime of the cosmological black hole can be observed in the frame  moving with a velocity $v$ with respect to the locally static observer  in the Kerr-dS spacetime. According to the Lorentz transformation formula in special relativity, the 3-velocity of the moving reference satisfies
\begin{equation}\label{LT}
  {\hat U}_{stat} \cdot {\hat U}_{mov} = - \frac{1}{\sqrt{1 - v^2}},
\end{equation}
where $\hat{U}$'s denote the four-velocities of the observers. From (\ref{et}), we can assign the four-velocity of the locally static observer in the Kerr-dS spacetime to be the normalized timelike vector $\hat{e}_{(t)}$, viz.,
\begin{equation}\label{Ustat}
  \hat{U}_{stat} = \sqrt{\frac{g_{\phi \phi}}{g_{t \phi}^2 - g_{t t} g_{\phi
  \phi}}} \left( \partial_t - \frac{g_{t \phi}}{g_{\phi \phi}} \partial_{\phi}\right).
\end{equation}
 As we mentioned before, the locally static observer have to stay in the {\em domain of outer communication}, however, as we will show in the following, there is no such restriction for the comoving observer in an expanding universe.

Next we need to find the four-velocity $U_{mov}$ of the comoving observer at large distance. First of all, at large distance, i.e. $r \gg \mathrm{max}(m, a,\Lambda)$, the
Kerr-dS metric becomes
\begin{eqnarray}\label{largerlimit}
  d s^2 & = & - \frac{\left( 1 - H_0^2 a^2 \sin^2 \theta -
  H_0^2  r^2 \right)}{\left( 1 + a^2 H_0^2  \right)^2
  \text{}} d t^2 + \frac{d r^2}{\left(1-H_0^2  a^2 -
  H_0^2 r^2 \right)} \nonumber\\
  &  & + \frac{r^2 d \theta^2}{1+H_0^2  a^2 \cos^2 \theta} +
  \frac{r^2 \sin^2 \theta d \phi^2}{\left(1+a^2 H_0^2
  \right)^{} \text{}} - \frac{2 a H_0^2  \sin^2 \theta r^2}{ \left( 1 +
  a^2 H_0^2  \right)^2 \text{}} d t d \phi.
\end{eqnarray}
Note that in order not to miss the nontrivial effect from the cosmological constant, in the above formula the subleading terms in the large $r$ limit have been kept. For example, if $H_0=0$, this expression correctly reproduces the Minkowski metric and let $a=0$ then it gives the de Sitter metric. Therefore, (\ref{largerlimit}) allows us to find the four-velocities of observers in the two separated regions within and beyond the {\em domain of outer communication}. For the locally static observer, the large distance limit has to be taken within the {\em domain of outer communication}, which requires that $r_C\gg r_+$. In this case, $H_0^2 a^2\ll1$ and (\ref{largerlimit}) is further simplified to be the de Sitter metric. So the four-velocity of the locally static observer at large distance becomes
\be\label{Ustatlarger}
\hat{U}_{stat} = \frac{1}{\sqrt{1-H_0^2r^2}} \left( \partial_t +a H_0^2  \partial_{\phi}\right).
\ee
Another notable thing is that in the large $r$ limit, the massless limit of the Kerr-dS metric (\ref{KerrdS_massless}) behaves exactly the same as the above formula, which means that (\ref{largerlimit}) can be related to the de Sitter metric and further be transformed into the form of the FLRW metric (\ref{FLRW}) in the large distance limit. So finally we build the connection between the Kerr-dS metric and the FLRW metric at least in the large distance limit. Naturally we can define the four-velocity of the comoving observer at large distance as
\begin{equation}
  \hat{U}_{mov} = N \partial_{\bar{t}},
\end{equation}
where $N$ is a normalization constant coefficient such that $\hat{U}_{mov}^2=-1$.
From the coordinate transformations (\ref{ptbar}), (\ref{Tf_KerrdS_m_0a}) and (\ref{Tf_KerrdS_m_0d}), in terms of the BL coordinates $ \hat{U}_{mov}$ can be expressed as
\begin{equation}\label{Umov}
  \hat{U}_{mov} = \frac{(1+a^2 H_0^2)^2}{1+a^2 H_0^2 \cos ^2\theta - H_0^2 r^2 \left(1+a^2H_0^2 \cos ^2\theta \right)} \partial_t
   + H_0 r \partial_r.
\end{equation}
Clearly, in contrast to the locally static observer, the comoving observer can locate beyond the {\em domain of outer communication}. Although according to the relation (\ref{LT}), the locally static observer cannot exceed the {\em domain of outer communication}, the comoving observer of interest here is subject to this constraint as well. However, we would like to stress that the above formula and the final formula for the shadow as seen by a
comoving observer in the following is not limited in the {\em domain of outer communication}, since it can be safely applied for the region
beyond the cosmological horizon too. Similar situation occurs for the comoving observers in the McVittie spacetime  \cite{Perlick:2018iye}. In this case, the four-velocity of the
comoving observer simplifies as
 \be\label{Umovlarger}
  U_{mov} = \frac{1}{1- H_0^2 r^2}\partial_t
   + H_0 r \partial_r,
 \ee
which is the same as the one for comoving observer in the McVittie spacetime at large distance \cite{Perlick:2018iye}.

Plugging (\ref{Ustatlarger}) and (\ref{Umovlarger}) into (\ref{LT}) we obtain
\begin{equation}
  - \frac{1}{\sqrt{1 - v^2}} = - \frac{1}{\sqrt{1 - H_0^2
  r^2}},
\end{equation}
which gives
\begin{equation}
  v = H_0 r.
\end{equation}
As expected, this result is independent of the  rotation of the black hole, since the large distance limit is taken in such a way that the spacetime is described by the de Sitter metric.
\begin{figure}[h]
\begin{center}
\includegraphics[width=75mm,angle=0]{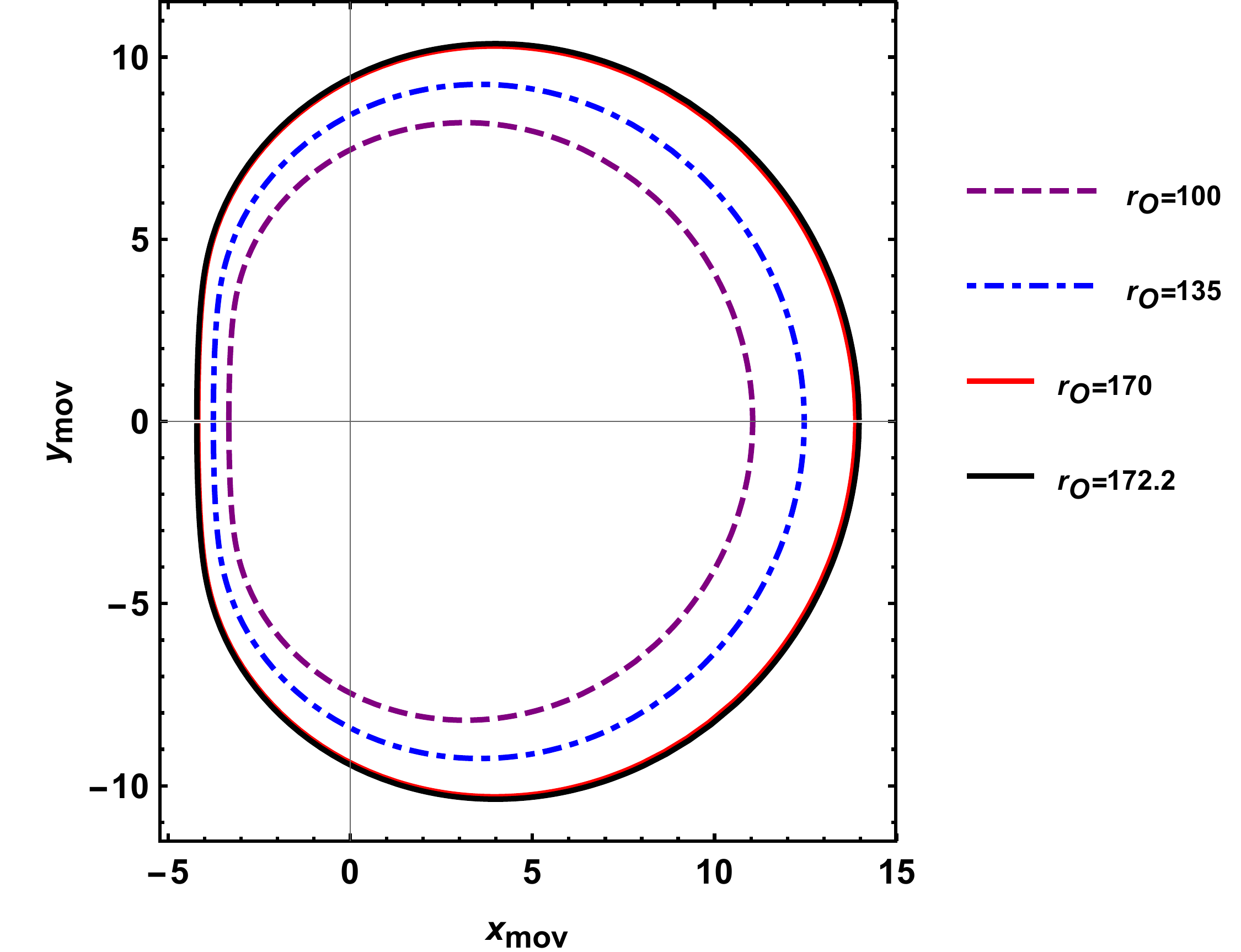}
\end{center}
\vspace{-5mm}
 \caption { Plot showing the shadows of a spinning black hole as seen by a comoving observer with different distances.
 The parameters are taken as $m=1, a=0.998, \Lambda=10^{-4}$ and $\theta_O=\pi/2$ }\label{shadowcom}
\end{figure}

 With the relation (\ref{LT}) between the locally static and comoving observers, we can find the angular size $\theta_{mov}$ of the black hole shadow seen by a comoving observer. The angular size $\theta_{stat}$ of the black hole shadow for the locally static observer has been calculated in the preceding section. Then $\theta_{mov}$ can be obtained by using the standard aberration formula
 \be\label{abformula}
 \sin^2\theta_{mov}=(1-v^2)\frac{ \sin^2\theta_{stat}}{(1-v \cos\theta_{stat})^2}.
 \ee
To use this formula we should first know the behavior of the solid angles $(\alpha, \beta)$  at large distance within the {\em domain of outer communication}. From (\ref{sima}) and (\ref{simb}) we  obtain the solid angles in the comoving observer's frame
\be
\sin \alpha_{mov}=\pm\frac{(1+H_0 r_O)\sqrt{\eta+(a-\xi )^2 -(\xi \csc \theta_O-a \sin \theta_O)^2}}{r_O}
,
\ee
and
\be
\sin \beta_{mov}=\frac{\xi (1+H_0 r_O) }
    { r_O \sin\theta_O}.
\ee
Fig. \ref{shadowcom} shows the shadows of a spinning black hole embedding in an expanding universe driven by the positive cosmological constant as seen by a comoving observer with different distances. One can see that the black hole size would not becomes vanishing when the observer is moving away from the black hole. This is very different from the locally static observer.  It is essentially a magnification effect of the cosmic expansion on the shadow size.

Of course the aberration formula applies to the horizontal and vertical angular diameters of the
shadow as well. That is,
\be\label{comovingsolidanglesa}
\sin \rho_v=\frac{(1+H_0 r_O)\sqrt{\eta+(a-\xi )^2 -(\xi \csc \theta_O-a \sin \theta_O)^2}}{r_O}\Big|_{r=r_{v}}
,
\ee
and
\be\label{comovingsolidanglesb}
\sin \rho_{h1}=\frac{\xi (1+H_0 r_O) }
    { r_O \sin\theta_O}\Big|_{r=r_{h1}},\quad\sin \rho_{h2}=-\frac{\xi (1+H_0 r_O) }
    { r_O \sin\theta_O}\Big|_{r=r_{h2}}.
\ee
From above formulas, we find that for a distant comoving observer, the angular size goes to a non-zero value since the comoving observer can live beyond the {\em domain of outer communication}. Specifically, from (\ref{shadowSchdS}) and (\ref{abformula}) we obtain the angular size of the shadow for distant observers in the case of $a=0$, that is
 \be
 \sin \rho_{h1}=\sin \rho_{h2}=3\sqrt{3}H_0 m+ \frac{3\sqrt{3}m\sqrt{1-27H_0^2m^2}}{r_O},
 \ee
 which agrees with the result of distant comoving observers in the McVittie spacetime \cite{Perlick:2018iye}.
 Also, we can see that the amplification effect due to the cosmic expansion acts equally on the two characteristic angular sizes of the shadow. It's useful to introduce two new parameters, i.e. the difference between the horizontal and vertical angular diameters of the shadow
\be
D_{vh}\equiv \left|2\rho_v-(\rho_{h1}+\rho_{h2})\right|,
\ee
and the oblateness \footnote{We would like to thank the anonymous referee for helping us clarify this point.}
\be
\delta_{vh}\equiv \frac{2\rho_v}{\rho_{h1}+\rho_{h2}}.
\ee
As a result, from Eqs. (\ref{comovingsolidanglesa}) and (\ref{comovingsolidanglesb}) we find that
\bea
D_{vh}&=&(1+H_0r_O)D_{vh}^0,\\
\delta_{vh}&=&\delta_{vh}^0,
\eea
for a distant observer, where $D_{vh}^0$ and $\delta_{vh}^0$ denote the difference and oblateness respectively, when the cosmological constant is vanishing. These results tell us the difference may become large when the observer is far away from the black hole, while the shape of the shadow remains unchanged. In the next section we will see this point generally.

\section{Shadow of a spinning black hole in an expanding universe\label{section4}}

In the previous section, we only considered the spinning black hole shadow in an expanding universe driven by the positive cosmological constant. In this section we extend the study to an expanding multi-component universe (with matter, radiation and dark energy). We will employ the approximate method proposed in \cite{Bisnovatyi-Kogan:2018vxl} to calculate the shadow of a Kerr black hole in the expanding FLRW universe as seen by a comoving observer \footnote{In a subsequent paper \cite{Tsupko:2019mfo}, the authors of \cite{Bisnovatyi-Kogan:2018vxl} developed a new method to calculate analytically the angular size of black hole shadow in the McVittie metric as seen by a comoving observer. The result of that paper is valid within the entire range of possible positions of observer, not only for large distance. }. This approximate method applies to the situation that the observer is very far from the black hole, the cosmic expansion is perceived only at very large scales and can be neglected near the black hole. The approximation is essentially the same as we used in the preceding section, so we expect that the two results are equal when the cosmic expansion is driven by the positive cosmological constant.

We now describe the approximate method adopted in \cite{Bisnovatyi-Kogan:2018vxl}. First,  let us introduce a radial coordinate $r_I$ which obeys
\begin{equation}\label{rIrO}
  r_+ \ll r_I \ll r_O .
\end{equation}
Clearly, $r_I$ bears two properties: one is that it is far enough away from the spinning
black hole such that the gravity of the black hole can be neglected, the other one is that
$r_I$ is small enough in comparison with the observer's position such that the effect of the
cosmic expansion can be neglected. Therefore, in the region $r\leq r_I$, we just need to calculate the shadow size of
the Kerr black hole and find the asymptotic form at large distance. Then in the region
 $r>r_I$, the effect of the cosmic expansion has to be taken into account, which is implemented by applying the angular size redshift relation to the shadow size.

The shadow of a Kerr black hole for the locally static observer was calculated in section \ref{section2}.
From Eqs. (\ref{sima}) and (\ref{simb}) with $\Lambda=0$, we know the solid angles in the frame of the observer located at ($r_I$, $\theta_I$) are given by
\begin{equation}\label{Kerrangluarsize}
  \sin \alpha \simeq \pm \frac{\sqrt{\eta + a^2 \cos^2 \theta_I -
  \xi^2 \cot^2 \theta_I}}{r_I}, \quad \tan \beta \simeq
  \frac{\xi}{\sin^{} \theta_I\, r_I^{}} .
\end{equation}
As we have known, since the shape of the Kerr black hole is not a perfect circle, we can use the
horizontal and vertical angular diameters of the shadow to characterize its angular size. The
horizontal angular diameter $\Delta h$ (\ref{horizontalang}) is determined by the zeroes of
$\alpha$ and the vertical angular diameter $\Delta v$ (\ref{verticalang}) corresponds to the position at which $\alpha$ takes a maximum. As we mentioned before, in general  for a given $\theta_I$ they can only be obtained numerically. In the following we just show the result for $\theta_I=\pi/2$ in which case the
horizontal and vertical angular diameters of the shadow can be analytically found. For $\theta_I=\pi/2$ the zeroes of $\alpha$ just correspond to $\eta=0$ and the solution to which is given by (\ref{rpp}),
so we have
\be
\sin \rho_{h1}=\frac{\xi}{ r_I^{}}\Big|_{r=r_{p-}},\quad \sin \rho_{h2}=-\frac{\xi}{ r_I^{}}\Big|_{r=r_{p+}},
\ee
where the explicit expression of $\xi$ is given in (\ref{xi}) and we have taken into account that for large $r_I$, $\sin\beta\simeq \tan\beta$. Moreover, for $\theta_I=\pi/2$, from Eq. (\ref{rofalphamax}) with $\Lambda=0$, it is easy to see that for the observer located at $r_I$, $\alpha$ takes maximum at $r=3m$. Then the vertical angular radius of the shadow is
\be
\sin \rho_v=\frac{\sqrt{\eta }}{r_I}\Big|_{r=3m}=\frac{\sqrt{27}m}{r_I}.
\ee
Note that this result is the same as the angular radius of the Schwarzschild black hole shadow.

Furthermore, we can formally define the linear size of the black hole shadow as seen by the observer at radius $r_I$  as
\be\label{linearsize}
R_h=r_I \sin \frac{\Delta h}{2}\simeq\frac{\xi}{2\sin^{} \theta_I\, }\Big|_{r=r_{h1}}-\frac{\xi}{2\sin^{} \theta_I\, }\Big|_{r=r_{h2}},
\quad R_v=\sqrt{\eta + a^2 \cos^2 \theta_I -
  \xi^2 \cot^2 \theta_I}\Big|_{r=r_{v}}.
\ee
Obviously, the above formulas are independent of $r_I$. In the non-rotating case, one has
$R_h=R_v=3\sqrt{3}m$.

In the region $r>r_I$, a distant observer located at $r_O$ finds the image of the black hole shadow with linear size characterized by $R_h$ and $R_v$ as well. But in the meantime, the null geodesic extending from the black hole out to the observer suffers from the cosmic expansion. As a result  the shadow size measured by the observer on earth should be different. The angular size redshift formula relates the linear size measured without the cosmic expansion to the angular size of the object as viewed by the observer, thus can be directly used for the shadow. Let the observer located on earth be the position with zero redshift, then the shadow of the black hole is situated at the redshift $z$.
The angular size redshift formula is not valid for the black hole shadow if the observer is close to the black hole such that the gravity of the black hole is dominant than the cosmic expansion. This is why the relation (\ref{rIrO}) is necessary.

According to the definition, the angular diameter distance is given by
\begin{equation}
  d_A = \frac{l}{\Delta \theta},
\end{equation}
where $l$ is the physical size of the object, and $\Delta \theta$ is the angular size of the object as viewed from earth.
For an object at the redshift $z$, the angular diameter distance can be expressed in terms of the comoving distance $\chi$ as
\begin{equation}
  d_A (z) = \frac{S_k(\chi)}{(1 + z) }.
\end{equation}
where $S_k(\chi)$ is the FLRW coordinate. When the universe is flat, we have $S_k(\chi)=\chi$ and the comoving distance is explicitly given by
\begin{equation}\label{chiz}
  \chi(z) = \int_0^z \frac{d z'}{H_0\sqrt{\Omega_{m 0} (1 + z')^3 + \Omega_{r 0} (1 + z')^4 +
  \Omega_{\Lambda 0}}} ,
\end{equation}
where $H_0$ is the present day value of the Hubble parameter $H (t)$, and $\Omega_{m
0}$, $\Omega_{r 0}$, $\Omega_{\Lambda 0}$ are the present day values of
the density parameters for matter, radiation and dark energy respectly. So
the angular size of the black hole shadow as seen by the observer on earth can be expressed as a function of the redshift
\begin{equation}
  \Delta \theta =  \frac{l  (1 + z)}{\chi(z)}.
\end{equation}
In this formula, the physical size $l$ corresponds to the horizontal and vertical linear sizes of the black hole shadow $R_h$ and $R_v$, so  we have respectively
\begin{equation}\label{hzvz}
  \Delta \theta_{h} (z) = R_{h}\frac{ (1 + z)}{\chi(z)},\quad \Delta\theta_{v} (z) = R_{v}\frac{(1 + z)}{\chi(z)}.
\end{equation}
Here for the observer on earth the angle $\theta_I$ should be replaced by the inclination angle $\theta_O$ between  the observer and the direction of the rotation axis of the black hole.
This formula allows to calculate the size of the shadow as a function of the redshift of the observed
black hole with given parameters $H_0$, $\Omega_{m 0}$,
$\Omega_{r 0}$, $\Omega_{\Lambda 0}$.

\begin{figure}[h]
\begin{center}
\includegraphics[width=120mm,angle=0]{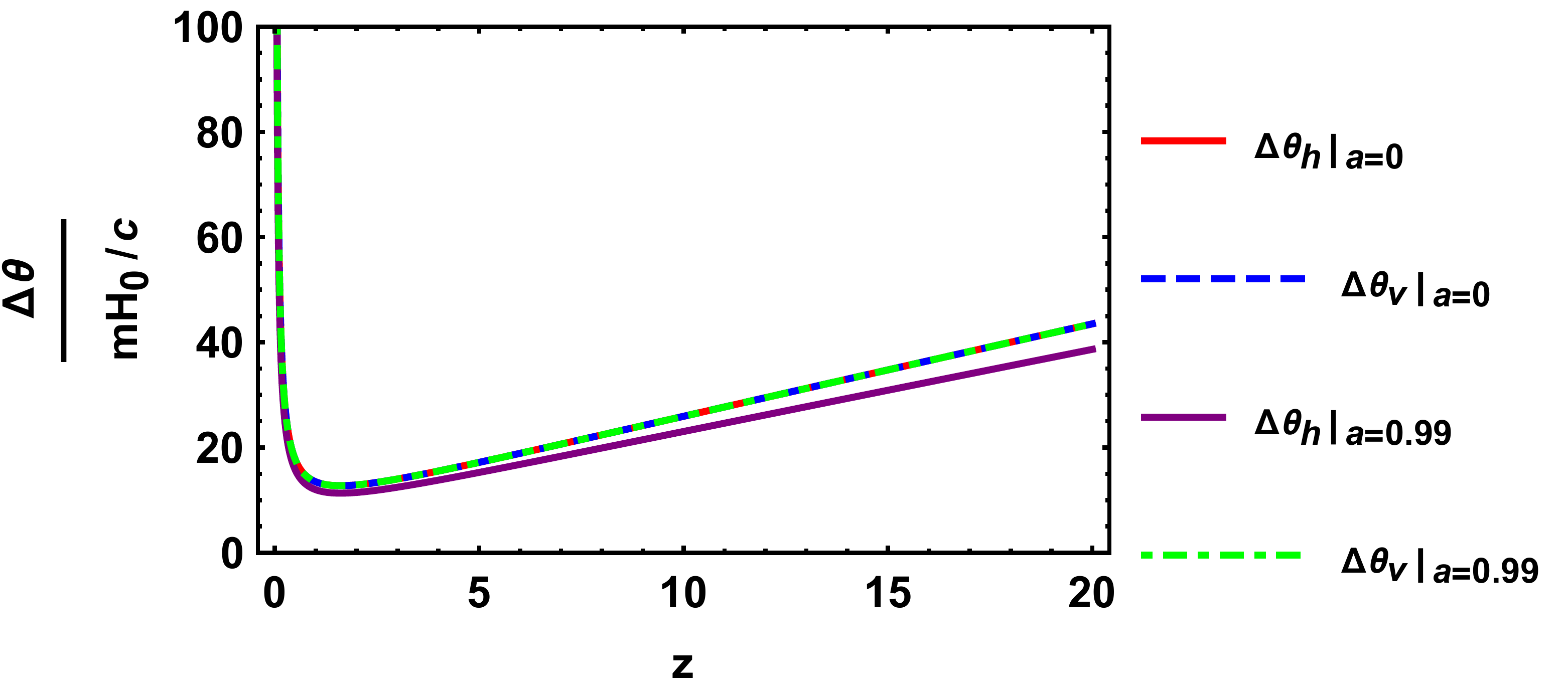}
\end{center}
\vspace{-5mm}
 \caption {The change of $\Delta\theta_h$ and $\Delta\theta_v$ with respect to the rotating parameter $a$, by choosing $\theta_O=\frac{\pi}{2}$.}\label{a0vsa99}
\end{figure}

\paragraph{De Sitter universe} First of all, let us consider that the expanding universe is totally driven by the cosmological constant.
For such de Sitter universe, we have $\Omega_{\Lambda0}=1$ and $\chi(z)=z/H_0$. Therefore, the angular sizes of the black hole  as seen by a comoving observer are given by
\be\label{linearsizecom}
 \Delta \theta_{h} (z) = R_{h}\frac{ H_0 (1 + z)}{z},\quad \Delta\theta_{v} (z) = R_{v}\frac{ H_0 (1 + z)}{z}.
\ee

\begin{figure}[h]
\begin{center}
\includegraphics[width=130mm,angle=0]{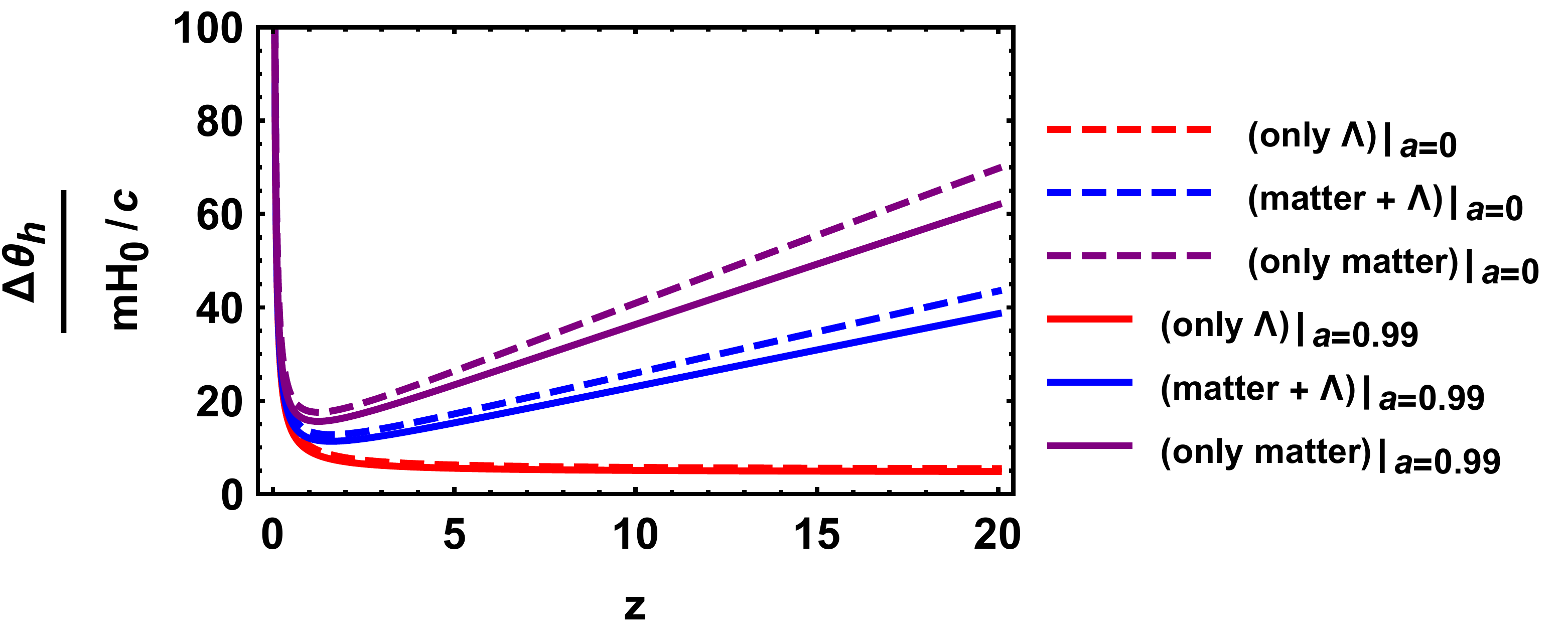}
\end{center}
\vspace{-5mm}
 \caption { The change of $\Delta\theta_h$ with respect to the rotating parameter $a$ in de Sitter universe, the matter-dominated  universe and the real multicomponent universe respectively, by choosing $\theta_O=\frac{\pi}{2}$. }\label{a0vsa991}
\end{figure}

Remember that in section \ref{section3} we have calculated the shadow of the spinning black hole embedded in the de Sitter universe as seen by a comoving observer at large distance. As we will show in the following, these two different methods lead to the same result. To prove that, a key step is to relate $r_O$ in Eqs. (\ref{comovingsolidanglesa}) and
(\ref{comovingsolidanglesb}) with the redshift $z$. At large distance $r_O\gg r_+$ and for small $H_0$, we know the relation between the static coordinate $r_O$ and the comoving coordinate $\bar{r}$
\be
r_O=\bar{r}\, e^{H_0 \bar{t}}.
\ee
Moreover, the comoving coordinate and the redshift has the following connection
\be
\bar{r}= \frac{1}{e^{H_0 \bar{t}}}\frac{z}{H_0}.
\ee
From these two formulas we can easily find that
\be\label{relambda}
\frac{1+H_O r_O}{r_O}=\frac{ H_0 (1 + z)}{z}.
\ee
Then compare Eqs. (\ref{comovingsolidanglesa}) and
(\ref{comovingsolidanglesb}) with (\ref{linearsizecom}), one can find that they are identical.

\paragraph{Matter-dominated  universe} In this universe, we have $\Omega_{m0}=1$, $\Omega_{r0}=0$ and $\Omega_{\Lambda0}=0$.
Then from (\ref{chiz}) and (\ref{hzvz}) we obtain
\begin{equation}\label{rematter}
  \Delta \theta_{h} (z) = R_{h}\frac{ (1 + z)}{2[1-(1+z)^{-1/2}]},\quad \Delta\theta_{v} (z) = R_{v}\frac{ (1 + z)}{2[1-(1+z)^{-1/2}]}.
\end{equation}
As found in \cite{Bisnovatyi-Kogan:2018vxl}, in this universe the angular size increases unboundedly with increase of the black hole redshift.

\paragraph{Real multicomponent universe}

The formula (\ref{hzvz}) allows us to calculate the angular size of the black hole shadow in the real universe with multiple components. For example we choose $\Omega_{\Lambda0}=0.7$ and $\Omega_{m0}=0.3$ with vanishing $\Omega_{r0}$, then the
pictures in this case are plotted in Figs. \ref{a0vsa99} and \ref{a0vsa991}.
Since the cosmic expansion appears as a multiplicative factor in the formula (\ref{hzvz}), the influence of different spin and inclination angle on the angular radii will be the same as that for a locally static observer, which has been clearly discussed in section \ref{section2}.

From Fig. \ref{a0vsa99}, one can see that with the increase of the redshift the difference between the horizontal radius and the vertical one caused by the high spin of the black hole  becomes significant and the inclination is $\pi/2$. In other words, the size of the black hole caused by the spin is magnified by the redshift. However, the ratio of the two radii keeps invariant, due to the fact that  the magnification effect of the cosmic expansion acts on them equally as can be seen from (\ref{hzvz}). Therefore the shape and the oblateness of the deformed shadow of a spinning black hole are unchanged with respect to the cosmic expansion. In addition, it might be possible to extract the information of the spin of the black hole from the observation of its shadow's deformation, as proposed by Tsupko in \cite{Tsupko:2017rdo}. Our study shows that it might be easier to achieve this goal for supermassive black holes at high redshift as long as the resolution of EHT is high enough. This is because the measurement of the angular radii would be easier due to the amplification of the cosmic expansion, so the accuracy of the oblateness of the shadow would be improved as well. Of course, this is purely a theoretical point of view. In realistic universe, the observation of shadows of high redshift supermassive black holes  is far beyond the ability of the current technology, see the concerns in \cite{Vagnozzi:2020quf}.

 In Fig. \ref{a0vsa991} we show the effect of the comic expansion caused by different components  on the horizontal radius of the shadow. One can see that the difference between the high spin one and the non-spinning one becomes the most significant when the component is purely matter.

\section{Summary}\label{summary}

In this paper, we studied the effect of the cosmic expansion on the shadow of the spinning black hole.  We started from the Kerr-dS spacetime. We may still define a distant observer comoving with the cosmic expansion  in the Kerr-dS spacetime. Using the usual method of calculating the shadow of a stationary black hole as seen by a locally static observer and the standard aberration formula, we  obtained the shadow of the Kerr black hole for a comoving observer, see  Eqs. (\ref{comovingsolidanglesa}) and (\ref{comovingsolidanglesb}). We found the shadow of the Kerr black hole seen by a locally static observer is very different from the one seen by a comoving observer. As for the former, the angular size of the Kerr-dS shadow becomes smaller as the observer moves further away from the black hole. Thus for a distant static observer, the angular size goes to be invisible, while for the comoving observer, the angular size of the shadow tends to a
constant. The existence of this difference gives us a reason to believe that this is indeed the size of the Kerr black hole shadow for a comoving observer with the cosmic expansion.

 Furthermore, we  considered the problem in the general expanding FLRW universe,  beyond the $\Lambda$-driven expansion. We applied the approximate method proposed by \cite{Bisnovatyi-Kogan:2018vxl}. We first concentrated on the $\Lambda$-driven case, and found the consistent result  with the one from the the direct computation. Then we turned to multi-component case.
 The main results are presented in Eq. (\ref{hzvz}),(\ref{linearsizecom}) and (\ref{rematter}), and the corresponding graph in Fig. \ref{a0vsa991}. The results show that the difference between the horizontal and vertical radii of the black hole shadow becomes significant in the case that the supermassive black hole is at a high redshift, but the shape is unaltered.

We conclude this paper with some outlooks. First of all, in our paper, we did not  include the influences of complex environments outside the black holes, such as the plasma, the jets or the accretion disc. It is definitely very interesting to consider these effects in the future works, since the light rays are usually refracted by the medium before they reach our eyes. Secondly, in the propagation of lights, the photon fields may interact with other fields, like axion etc. which is also worth considering in future. In the end,  we have been focusing on the shape and the size of a black hole shadow in this work. Another important observable, the brightness of edges of the shadows, has not been studied here. As we know, in the EHT experiment, the factor of luminosity is  crucial as well. It should be considered carefully in further studies.

\section*{Acknowledgments}
We thank Zhong-Ying Fan, Canbin Liang and Xiaoning Wu for useful discussions. The work is in part supported by NSFC Grant No. 11335012, No. 11325522, No. 11735001 and No. 11847241. MG and PCL are also supported by NSFC Grant No. 11947210. And MG is also funded by China National Postdoctoral Innovation Program 2019M660278.


\begin{thebibliography}{10}

\bibitem{Akiyama:2019cqa}
  K.~Akiyama {\it et al.} [Event Horizon Telescope Collaboration],
  ``First M87 Event Horizon Telescope Results. I. The Shadow of the Supermassive Black Hole,''
  Astrophys.\ J.\  {\bf 875}, no. 1, L1 (2019)
  doi:10.3847/2041-8213/ab0ec7
  [arXiv:1906.11238 [astro-ph.GA]].

 \bibitem{Akiyama:2019brx}
K.~Akiyama {\it et al.} [Event Horizon Telescope Collaboration],
  ``First M87 Event Horizon Telescope Results. II. Array and Instrumentation,''
  Astrophys.\ J.\  {\bf 875}, no. 1, L2 (2019)
  doi:10.3847/2041-8213/ab0c96
  [arXiv:1906.11239 [astro-ph.IM]].

  \bibitem{Akiyama:2019sww}
  K.~Akiyama {\it et al.} [Event Horizon Telescope Collaboration],
  ``First M87 Event Horizon Telescope Results. III. Data Processing and Calibration,''
  Astrophys.\ J.\  {\bf 875}, no. 1, L3 (2019)
  doi:10.3847/2041-8213/ab0c57
  [arXiv:1906.11240 [astro-ph.GA]].

  \bibitem{Akiyama:2019bqs}
  K.~Akiyama {\it et al.} [Event Horizon Telescope Collaboration],
  ``First M87 Event Horizon Telescope Results. IV. Imaging the Central Supermassive Black Hole,''
  Astrophys.\ J.\  {\bf 875}, no. 1, L4 (2019)
  doi:10.3847/2041-8213/ab0e85
  [arXiv:1906.11241 [astro-ph.GA]].

  \bibitem{Akiyama:2019fyp}
  K.~Akiyama {\it et al.} [Event Horizon Telescope Collaboration],
  ``First M87 Event Horizon Telescope Results. V. Physical Origin of the Asymmetric Ring,''
  Astrophys.\ J.\  {\bf 875}, no. 1, L5 (2019)
  doi:10.3847/2041-8213/ab0f43
  [arXiv:1906.11242 [astro-ph.GA]].

  \bibitem{Akiyama:2019eap}
  K.~Akiyama {\it et al.} [Event Horizon Telescope Collaboration],
  ``First M87 Event Horizon Telescope Results. VI. The Shadow and Mass of the Central Black Hole,''
  Astrophys.\ J.\  {\bf 875}, no. 1, L6 (2019)
  doi:10.3847/2041-8213/ab1141
  [arXiv:1906.11243 [astro-ph.GA]].

  \bibitem{Synge:1966okc}
  J.~L.~Synge,
  ``The Escape of Photons from Gravitationally Intense Stars,''
  Mon.\ Not.\ Roy.\ Astron.\ Soc.\  {\bf 131}, no. 3, 463 (1966).

 \bibitem{Bardeen:1973}
  J. M. Bardeen, ``Timelike and null geodesies in the Kerr metric," in {\em Proceedings, Ecole
d'Ete de Physique Theorique: Les Astres Occlus}, eds. C. Witt and B. Witt (Les
Houches, France, 1973), p. 215-239.

 \bibitem{Chandrasekhar:1992}
S. Chandrasekhar, {\em The Mathematical Theory of Black Holes} (Oxford University Press, New York, 1992).

 \bibitem{Cunningham:1972}
C. Cunningham and J. Bardeen, ``The optical
appearance of a star orbiting an extreme kerr black
hole,'' The Astrophysical Journal {\bf173} (1972) L137.

 \bibitem{Cunningham:1973}
C. Cunningham and J. M. Bardeen, ``The optical
appearance of a star orbiting an extreme kerr black
hole,'' The Astrophysical Journal {\bf183} (1973) 237-264.

\bibitem{Porfyriadis:2016gwb}
  A.~P.~Porfyriadis, Y.~Shi and A.~Strominger,
  ``Photon Emission Near Extreme Kerr Black Holes,''
  Phys.\ Rev.\ D {\bf 95}, no. 6, 064009 (2017)
  doi:10.1103/PhysRevD.95.064009
  [arXiv:1607.06028 [gr-qc]].

\bibitem{Gralla:2017ufe}
  S.~E.~Gralla, A.~Lupsasca and A.~Strominger,
  ``Observational Signature of High Spin at the Event Horizon Telescope,''
  Mon.\ Not.\ Roy.\ Astron.\ Soc.\  {\bf 475}, no. 3, 3829 (2018)
  doi:10.1093/mnras/sty039
  [arXiv:1710.11112 [astro-ph.HE]].

  \bibitem{Guo:2018kis}
  M.~Guo, N.~A.~Obers and H.~Yan,
  ``Observational signatures of near-extremal Kerr-like black holes in a modified gravity theory at the Event Horizon Telescope,''
  Phys.\ Rev.\ D {\bf 98}, no. 8, 084063 (2018)
  doi:10.1103/PhysRevD.98.084063
  [arXiv:1806.05249 [gr-qc]].

\bibitem{Gates:2018hub}
  D.~Gates, D.~Kapec, A.~Lupsasca, Y.~Shi and A.~Strominger,
  ``Polarization Whorls from M87 at the Event Horizon Telescope,''
  arXiv:1809.09092 [hep-th].

  \bibitem{Long:2018tij}
  F.~Long, S.~Chen, J.~Wang and J.~Jing,
  ``Electromagnetic emissions from near-horizon region of an extreme Kerr-Taub-Nut black hole,''
  Eur.\ Phys.\ J.\ C {\bf 79}, no. 6, 466 (2019)
  doi:10.1140/epjc/s10052-019-6989-8
  [arXiv:1812.11463 [gr-qc]].

\bibitem{Yan:2019etp}
  H.~Yan,
  ``Influence of a plasma on the observational signature of a high-spin Kerr black hole,''
  Phys.\ Rev.\ D {\bf 99}, no. 8, 084050 (2019)
  doi:10.1103/PhysRevD.99.084050
  [arXiv:1903.04382 [gr-qc]].

  \bibitem{Igata:2019pgb}
  T.~Igata, H.~Ishihara and Y.~Yasunishi,
  ``Observability of spherical photon orbits in near-extremal Kerr black holes,''
  Phys.\ Rev.\ D {\bf 100}, no. 4, 044058 (2019)
  doi:10.1103/PhysRevD.100.044058
  [arXiv:1904.00271 [gr-qc]].

  \bibitem{Kapec:2019hro}
  D.~Kapec and A.~Lupsasca,
  ``Particle motion near high-spin black holes,''
  Class.\ Quant.\ Grav.\  {\bf 37}, no. 1, 015006 (2020)
  doi:10.1088/1361-6382/ab519e
  [arXiv:1905.11406 [hep-th]].

  \bibitem{Igata:2019hkz}
  T.~Igata, K.~Nakashi and K.~Ogasawara,
  ``Observability of the innermost stable circular orbit in a near-extremal Kerr black hole,''
  arXiv:1910.12682 [astro-ph.HE].

  \bibitem{Guo:2019lur}
  M.~Guo, S.~Song and H.~Yan,
  ``Observational signature of a near-extremal Kerr-Sen black hole in the heterotic string theory,''
  arXiv:1911.04796 [gr-qc].

  \bibitem{Guo:2019pte}
  M.~Guo, P.~C.~Li and B.~Chen,
  ``Photon Emission Near Myers-Perry Black Holes in the Large Dimension Limit,''
  arXiv:1911.08814 [gr-qc].

  \bibitem{Grenzebach:2014fha}
  A.~Grenzebach, V.~Perlick and C.~Lämmerzahl,
  ``Photon Regions and Shadows of Kerr-Newman-NUT Black Holes with a Cosmological Constant,''
  Phys.\ Rev.\ D {\bf 89}, no. 12, 124004 (2014)
  doi:10.1103/PhysRevD.89.124004
  [arXiv:1403.5234 [gr-qc]].


\bibitem{Grenzebach:2015oea}
  A.~Grenzebach, V.~Perlick and C.~Lämmerzahl,
  ``Photon Regions and Shadows of Accelerated Black Holes,''
  Int.\ J.\ Mod.\ Phys.\ D {\bf 24}, no. 09, 1542024 (2015)
  doi:10.1142/S0218271815420249
  [arXiv:1503.03036 [gr-qc]].

\bibitem{Amir:2016cen}
  M.~Amir and S.~G.~Ghosh,
  ``Shapes of rotating nonsingular black hole shadows,''
  Phys.\ Rev.\ D {\bf 94}, no. 2, 024054 (2016)
  doi:10.1103/PhysRevD.94.024054
  [arXiv:1603.06382 [gr-qc]].

  \bibitem{Abdujabbarov:2016hnw}
  A.~Abdujabbarov, M.~Amir, B.~Ahmedov and S.~G.~Ghosh,
  ``Shadow of rotating regular black holes,''
  Phys.\ Rev.\ D {\bf 93}, no. 10, 104004 (2016)
  doi:10.1103/PhysRevD.93.104004
  [arXiv:1604.03809 [gr-qc]].



  \bibitem{Dastan:2016vhb}
  S.~Dastan, R.~Saffari and S.~Soroushfar,
  ``Shadow of a Charged Rotating Black Hole in $f(R)$ Gravity,''
  arXiv:1606.06994 [gr-qc].

  \bibitem{Younsi:2016azx}
  Z.~Younsi, A.~Zhidenko, L.~Rezzolla, R.~Konoplya and Y.~Mizuno,
  ``New method for shadow calculations: Application to parametrized axisymmetric black holes,''
  Phys.\ Rev.\ D {\bf 94}, no. 8, 084025 (2016)
  doi:10.1103/PhysRevD.94.084025
  [arXiv:1607.05767 [gr-qc]].

\bibitem{Wang:2017hjl}
  M.~Wang, S.~Chen and J.~Jing,
  ``Shadow casted by a Konoplya-Zhidenko rotating non-Kerr black hole,''
  JCAP {\bf 1710}, 051 (2017)
  doi:10.1088/1475-7516/2017/10/051
  [arXiv:1707.09451 [gr-qc]].

   \bibitem{Cunha:2018acu}
  P.~V.~P.~Cunha and C.~A.~R.~Herdeiro,
  ``Shadows and strong gravitational lensing: a brief review,''
  Gen.\ Rel.\ Grav.\  {\bf 50}, no. 4, 42 (2018)
  doi:10.1007/s10714-018-2361-9
  [arXiv:1801.00860 [gr-qc]].

  \bibitem{Wang:2018eui}
  M.~Wang, S.~Chen and J.~Jing,
  ``Chaotic shadow of a non-Kerr rotating compact object with quadrupole mass moment,''
  Phys.\ Rev.\ D {\bf 98}, no. 10, 104040 (2018)
  doi:10.1103/PhysRevD.98.104040
  [arXiv:1801.02118 [gr-qc]].

  \bibitem{Hennigar:2018hza}
  R.~A.~Hennigar, M.~B.~J.~Poshteh and R.~B.~Mann,
  ``Shadows, Signals, and Stability in Einsteinian Cubic Gravity,''
  Phys.\ Rev.\ D {\bf 97}, no. 6, 064041 (2018)
  doi:10.1103/PhysRevD.97.064041
  [arXiv:1801.03223 [gr-qc]].

  \bibitem{Ovgun:2018tua}
  A.~Övgün, İ.~Sakalli and J.~Saavedra,
  ``Shadow cast and Deflection angle of Kerr-Newman-Kasuya spacetime,''
  JCAP {\bf 1810}, 041 (2018)
  doi:10.1088/1475-7516/2018/10/041
  [arXiv:1807.00388 [gr-qc]].

  \bibitem{Haroon:2018ryd}
  S.~Haroon, M.~Jamil, K.~Jusufi, K.~Lin and R.~B.~Mann,
  ``Shadow and Deflection Angle of Rotating Black Holes in Perfect Fluid Dark Matter with a Cosmological Constant,''
  Phys.\ Rev.\ D {\bf 99}, no. 4, 044015 (2019)
  doi:10.1103/PhysRevD.99.044015
  [arXiv:1810.04103 [gr-qc]].

  \bibitem{Wang:2018prk}
  H.~M.~Wang, Y.~M.~Xu and S.~W.~Wei,
  ``Shadows of Kerr-like black holes in a modified gravity theory,''
  JCAP {\bf 1903}, 046 (2019)
  doi:10.1088/1475-7516/2019/03/046
  [arXiv:1810.12767 [gr-qc]].

  \bibitem{Wei:2019pjf}
  S.~W.~Wei, Y.~C.~Zou, Y.~X.~Liu and R.~B.~Mann,
  ``Curvature radius and Kerr black hole shadow,''
  JCAP {\bf 1908}, 030 (2019)
  doi:10.1088/1475-7516/2019/08/030
  [arXiv:1904.07710 [gr-qc]].

  \bibitem{Kumar:2019ohr}
  R.~Kumar, B.~P.~Singh and S.~G.~Ghosh,
  ``Rotating black hole shadow in asymptotically safe gravity,''
  arXiv:1904.07652 [gr-qc].

  \bibitem{Shaikh:2019fpu}
  R.~Shaikh,
  ``Black hole shadow in a general rotating spacetime obtained through Newman-Janis algorithm,''
  Phys.\ Rev.\ D {\bf 100}, no. 2, 024028 (2019)
  doi:10.1103/PhysRevD.100.024028
  [arXiv:1904.08322 [gr-qc]].

  \bibitem{Bambi:2019tjh}
  C.~Bambi, K.~Freese, S.~Vagnozzi and L.~Visinelli,
  ``Testing the rotational nature of the supermassive object M87* from the circularity and size of its first image,''
  Phys.\ Rev.\ D {\bf 100}, no. 4, 044057 (2019)
  doi:10.1103/PhysRevD.100.044057
  [arXiv:1904.12983 [gr-qc]].

  \bibitem{Konoplya:2019sns}
  R.~A.~Konoplya,
  ``Shadow of a black hole surrounded by dark matter,''
  Phys.\ Lett.\ B {\bf 795}, 1 (2019)
  doi:10.1016/j.physletb.2019.05.043
  [arXiv:1905.00064 [gr-qc]].

  \bibitem{Contreras:2019nih}
  E.~Contreras, J.~M.~Ramirez-Velasquez, Á.~Rincón, G.~Panotopoulos and P.~Bargueño,
  ``Black hole shadow of a rotating polytropic black hole by the Newman–Janis algorithm without complexification,''
  Eur.\ Phys.\ J.\ C {\bf 79}, no. 9, 802 (2019)
  doi:10.1140/epjc/s10052-019-7309-z
  [arXiv:1905.11443 [gr-qc]].

   \bibitem{Jusufi:2019nrn}
  K.~Jusufi, M.~Jamil, P.~Salucci, T.~Zhu and S.~Haroon,
  ``Black Hole Surrounded by a Dark Matter Halo in the M87 Galactic Center and its Identification with Shadow Images,''
  Phys.\ Rev.\ D {\bf 100}, no. 4, 044012 (2019)
  doi:10.1103/PhysRevD.100.044012
  [arXiv:1905.11803 [physics.gen-ph]].

  \bibitem{Vagnozzi:2019apd}
  S.~Vagnozzi and L.~Visinelli,
  ``Hunting for extra dimensions in the shadow of M87*,''
  Phys.\ Rev.\ D {\bf 100}, no. 2, 024020 (2019)
  doi:10.1103/PhysRevD.100.024020
  [arXiv:1905.12421 [gr-qc]].



  \bibitem{Zhu:2019ura}
  T.~Zhu, Q.~Wu, M.~Jamil and K.~Jusufi,
  ``Shadows and deflection angle of charged and slowly rotating black holes in Einstein-Æther theory,''
  Phys.\ Rev.\ D {\bf 100}, no. 4, 044055 (2019)
  doi:10.1103/PhysRevD.100.044055
  [arXiv:1906.05673 [gr-qc]].

  \bibitem{Ovgun:2019jdo}
  A.~Övgün, İ.~Sakalli, J.~Saavedra and C.~Leiva,
  ``Shadow cast of non-commutative black holes in Rastall gravity,''
  arXiv:1906.05954 [hep-th].

  \bibitem{Contreras:2019cmf}
  E.~Contreras, Á.~Rincón, G.~Panotopoulos, P.~Bargueño and B.~Koch,
  ``Black hole shadow of a rotating scale--dependent black hole,''
  arXiv:1906.06990 [gr-qc].

  \bibitem{Konoplya:2019goy}
  R.~A.~Konoplya and A.~Zhidenko,
  ``Analytical representation for metrics of scalarized Einstein-Maxwell black holes and their shadows,''
  Phys.\ Rev.\ D {\bf 100}, no. 4, 044015 (2019)
  doi:10.1103/PhysRevD.100.044015
  [arXiv:1907.05551 [gr-qc]].

\bibitem{Konoplya:2019fpy}
  R.~A.~Konoplya, T.~Pappas and A.~Zhidenko,
  ``Einstein--scalar--Gauss--Bonnet black holes: Analytical approximation for the metric and applications to calculations of shadows,''
  arXiv:1907.10112 [gr-qc].

  \bibitem{Das:2019sty}
  A.~Das, A.~Saha and S.~Gangopadhyay,
  ``Shadow of charged black holes in Gauss-Bonnet gravity,''
  arXiv:1909.01988 [gr-qc].

  \bibitem{Lu:2019zxb}
  H.~Lu and H.~D.~Lyu,
  ``On the Size of a Black Hole: The Schwarzschild is the Biggest,''
  arXiv:1911.02019 [gr-qc].

  \bibitem{Chang:2019vni}
  Z.~Chang and Q.~H.~Zhu,
  ``Black hole shadow in the view of freely falling observers,''
  arXiv:1911.02190 [gr-qc].

  \bibitem{Feng:2019zzn}
  X.~H.~Feng and H.~Lu,
  ``On the Size of Rotating Black Holes,''
  arXiv:1911.12368 [gr-qc].


  \bibitem{Kumar:2019pjp}
  R.~Kumar, S.~G.~Ghosh and A.~Wang,
  ``Shadow cast and deflection of light by charged rotating regular black holes,''
  Phys.\ Rev.\ D {\bf 100}, no. 12, 124024 (2019)
  doi:10.1103/PhysRevD.100.124024
  [arXiv:1912.05154 [gr-qc]].

  \bibitem{Ma:2019ybz}
  L.~Ma and H.~Lu,
  ``Bounds on Photon Spheres and Shadows of Charged Black Holes in Einstein-Gauss-Bonnet-Maxwell Gravity,''
  arXiv:1912.05569 [gr-qc].

    \bibitem{Allahyari:2019jqz}
  A.~Allahyari, M.~Khodadi, S.~Vagnozzi and D.~F.~Mota,
  ``Magnetically charged black holes from non-linear electrodynamics and the Event Horizon Telescope,''
  arXiv:1912.08231 [gr-qc].

  \bibitem{Kumar:2020hgm}
  R.~Kumar, S.~G.~Ghosh and A.~Wang,
  ``Light deflection and shadow cast by rotating Kalb-Ramond black holes,''
  arXiv:2001.00460 [gr-qc].

  \bibitem{Chang:2020miq}
  Z.~Chang and Q.~H.~Zhu,
  ``A revisit of rotating black hole shadow with astrometric observables,''
  arXiv:2001.05175 [gr-qc].

  \bibitem{Mishra:2019trb}
  A.~K.~Mishra, S.~Chakraborty and S.~Sarkar,
  ``Understanding photon sphere and black hole shadow in dynamically evolving spacetimes,''
  Phys.\ Rev.\ D {\bf 99}, no. 10, 104080 (2019)
  doi:10.1103/PhysRevD.99.104080
  [arXiv:1903.06376 [gr-qc]].

  \bibitem{Perlick:2018iye}
  V.~Perlick, O.~Y.~Tsupko and G.~S.~Bisnovatyi-Kogan,
  ``Black hole shadow in an expanding universe with a cosmological constant,''
  Phys.\ Rev.\ D {\bf 97}, no. 10, 104062 (2018)
  doi:10.1103/PhysRevD.97.104062
  [arXiv:1804.04898 [gr-qc]].

   \bibitem{Bisnovatyi-Kogan:2018vxl}
  G.~S.~Bisnovatyi-Kogan and O.~Y.~Tsupko,
  ``Shadow of a black hole at cosmological distances,''
  Phys.\ Rev.\ D {\bf 98}, no. 8, 084020 (2018)
  doi:10.1103/PhysRevD.98.084020
  [arXiv:1805.03311 [gr-qc]].

   \bibitem{Tsupko:2019mfo}
  O.~Y.~Tsupko and G.~S.~Bisnovatyi-Kogan,
  ``First analytical calculation of black hole shadow in McVittie metric,''
  arXiv:1912.07495 [gr-qc].

  \bibitem{Tsupko:2019pzg}
  O.~Y.~Tsupko, Z.~Fan and G.~S.~Bisnovatyi-Kogan,
  ``Black hole shadow as a standard ruler in cosmology,''
  arXiv:1905.10509 [gr-qc].

  \bibitem{Qi:2019zdk}
  J.~Z.~Qi and X.~Zhang,
  ``A new cosmological probe from supermassive black hole shadows,''
  arXiv:1906.10825 [astro-ph.CO].

  \bibitem{Vagnozzi:2020quf}
  S.~Vagnozzi, C.~Bambi and L.~Visinelli,
  ``Concerns regarding the use of black hole shadows as standard rulers,''
  arXiv:2001.02986 [gr-qc].



\bibitem{Akcay:2010vt}
  S.~Akcay and R.~A.~Matzner,
  ``Kerr-de Sitter Universe,''
  Class.\ Quant.\ Grav.\  {\bf 28}, 085012 (2011)
  doi:10.1088/0264-9381/28/8/085012
  [arXiv:1011.0479 [gr-qc]].

  \bibitem{Cunha:2016bpi}
  P.~V.~P.~Cunha, C.~A.~R.~Herdeiro, E.~Radu and H.~F.~Runarsson,
  ``Shadows of Kerr black holes with and without scalar hair,''
  Int.\ J.\ Mod.\ Phys.\ D {\bf 25}, no. 09, 1641021 (2016)
  doi:10.1142/S0218271816410212
  [arXiv:1605.08293 [gr-qc]].

  \bibitem{Stuchlik:1999qk}
  Z.~Stuchlik and S.~Hledik,
  ``Some properties of the Schwarzschild-de Sitter and Schwarzschild - anti-de Sitter space-times,''
  Phys.\ Rev.\ D {\bf 60}, 044006 (1999).

\bibitem{McVittie:1933}
 G. C. McVittie, ``The mass-particle in an expanding universe,"
 Mon. Not. Roy. Astron. Soc. 93, 325 (1933).


  \bibitem{Vaidya:1977zza}
  P.~C.~Vaidya,
  ``The Kerr metric in cosmological background,''
  Pramana {\bf 8}, 151 (1977).
  doi:10.1007/BF02872099

\bibitem{Thakurta:1981}
  S. N. G. Thakurta, ``Kerr metric in an expanding universe,"
Indian J. Phys. B {\bf55}, 304 (1981).

\bibitem{Garat:2000pn}
  A.~Garat and R.~H.~Price,
  ``Nonexistence of conformally flat slices of the Kerr space-time,''
  Phys.\ Rev.\ D {\bf 61}, 124011 (2000)
  doi:10.1103/PhysRevD.61.124011
  [gr-qc/0002013].

  \bibitem{Tsupko:2017rdo}
  O.~Y.~Tsupko,
  ``Analytical calculation of black hole spin using deformation of the shadow,''
  Phys.\ Rev.\ D {\bf 95}, no. 10, 104058 (2017)
  doi:10.1103/PhysRevD.95.104058
  [arXiv:1702.04005 [gr-qc]].

\end{thebibliography}
\end{document}